\documentclass[preprint,showpacs,preprintnumbers,amsmath,amssymb]{revtex4-1}
\usepackage{graphicx}
\usepackage{bm}
\usepackage{color}
\usepackage{amssymb}
\usepackage{epsfig}
\usepackage{graphicx}
\usepackage{rotating}
\begin{document}

\title{Spectral and Strength Statistics of Chiral Brownian Ensemble}
\author{Pragya Shukla}
\affiliation{ Department of Physics, Indian Institute of Technology, Kharagpur-721302, West Bengal, India }
\date{\today}

\widetext

\begin{abstract}

Multi-parametric chiral random matrix ensembles are important  tools to analyze the statistical behavior  of generic complex systems with chiral symmetry.  A recent study \cite{psmulti} of the former maps them to the chiral Brownian ensemble (Ch-BE) that  appears as a non-equilibrium state of a single parametric crossover between two stationary chiral Hermitian ensembles.  This motivates us to pursue a detailed statistical investigation of the spectral and strength fluctuations of the Ch-BE, with a focus on their behavior near zero 
energy region. The  information can then be used for a wide range of complex systems with chiral symmetry.   Our analysis also reveals  connections of Ch-BE to  generalized Calogero Sutherland Hamiltonian (CSH) and Wishart ensembles. This along with already known connections of complex systems without chirality to CSH strongly hints the later to be the  ''backbone'' Hamiltonian governing the spectral dynamics of Complex systems.

\end{abstract}

\maketitle
.

\section{Introduction}

Based on exact symmetries,  statistical behavior of  the eigenvalues and eigenfunctions in a complex system can be classified in various universality classes of the random matrix ensembles. A variation of the symmetry or an approximate symmetry breaking can lead to perturbation of the initial universality class, resulting in transition of the statistical behavior from one class to another. The need to understand the behavior during the transition motivated the introduction of Dyson's Brownian motion model which describes diffusion of the elements of a Hermitian matrix, with only anti-unitary symmetries, due to a random symmetry breaking perturbation; the ensemble of Hermitian matrices corresponding to intermediate state of diffusion are referred as Dyson's Brownian ensembles. Detailed numerical investigations have confirmed the success of these  ensembles as models for complex systems but only those with anti-unitary symmetries.  The latter however often  manifest  in company with other symmetries, thus making it necessary to seek similar models for systems with both  unitary as well as anti-unitary symmetries. An important case in this context is that of chiral symmetry which appears in many areas of current interest
 e.g. charge transport in graphene \cite{fm}, spectral fluctuations in QCD Dirac operators \cite{vb}, localization in bipartite lattices \cite{gade,gw,lfsg,sf,gll,mdh},  vortex glass problems \cite{co, td,hf}, conductance fluctuations in mesoscopic systems \cite{sn,zirn}, topological systems etc \cite{been-m, chal, smb, ek}. The objective of the present study is to consider a generalization of the Dyson's approach   to include chiral symmetry in presence of the anti-unitary symmetries; the corresponding ensembles are referred as the chiral Brownian ensembles.

For complex systems with ergodic/delocalized wave-dynamics, the combination of chiral symmetry  with  time-reversal as well rotational symmetry results in   three main universality classes of the statistical behavior.  
 Assuming independent and identical Gaussian randomness of the matrix elements, the chiral universality classes can briefly be described as the   chiral Gaussian orthogonal ensemble (Ch-GOE) for chiral real-symmetric matrix ensembles with time-reversal symmetry and integer angular momentum (labelled by a parameter $\beta=1$),   chiral Gaussian symplectic ensemble (Ch-GSE) for chiral real-quaternion matrix ensembles in presence of time-reversal symmetry and half-integer angular momentum ($\beta=4$),   chiral Gaussian unitary ensemble (Ch-GUE) for chiral complex Hermitian matrix ensembles in absence of time-reversal symmetry ($\beta=2$) \cite{vb}.   (As the  transformation properties of the matrix, representing the linear operator e.g. Hamiltonian, in the basis preserving both chirality as well time-reversal, are governed by the latter, with chirality manifesting only in the block structure, the chiral universality classes are named in analogy with their non-chiral  counterparts. For Gaussian distributions of the matrix elements, the latter correspond to  Gaussian ensembles of  Hermitian matrices invariant under orthogonal (O), unitary (U) or symplectic (S) transformations, also known as GOE, GUE, GSE or collectively referred as the Wigner-Dyson ensembles \cite{me, fh}. 
 
 Similar to Wigner-Dyson ensembles \cite{me,fh}, their chiral counterparts are  the ensemble of Hermitian matrices with independently and identically distributed (i.i.d) entries in off-diagonal blocks (with zero diagonal blocks) and are applicable for the chiral systems under local conditions which lead to almost all matrix elements of the same order. As the latter leaves the ensemble essentially free of any parameters  and invariant under change of basis, these ensembles are also referred as stationary or basis-invariant chiral ensembles. The basis-invariance of both chiral/ non-chiral ensembles manifests itself in form of their ergodic/delocalized eigenfunctions but an important difference appears in their spectrum.
The chiral symmetry induces an additional level repulsion around zero which for chiral ensembles results in  spectral correlations  near zero (the origin) different from those far from zero (the bulk). In the bulk however the spectral correlations are not affected by the block structure ensuring survival of the statistical behavior analogous to Winger-Dyson universality classes \cite{vb}.

Based on the nature of symmetry-breaking perturbation, say $V$, a stationary chiral ensemble, say $H_0$, can undergo many types of transitions. For example,  a random perturbation preserving chiral symmetry as well as basis-invariance of the ensemble of $H_0$ matrices  but breaking its anti-unitary symmetry results in transition between two stationary chiral ensembles; the intermediate states $H=H_0+ \lambda V$ of the transition are referred as chiral Brownian ensemble (with $\lambda$ as the perturbation parameter). In case $V$ breaks the chiral symmetry of $H_0$ while preserving its anti-unitarity as well as basis-invariance,  the transition then takes place from chiral universality class to Wigner-Dyson one, with non-equilibrium stages described by non-chiral Brownian ensembles (see for example \cite{kk,ms} for QCD-specific studies and \cite{kt} for particle-hole symmetric BDG Hamiltonians). However for cases when $V$ belongs to a basis-dependent ensemble, the intermediate states are then no longer Brownian ensembles. In the present work, we confine the analysis only to the chiral Brownian ensemble for arbitrary $H_0$ and $V$; (note a similar analysis, for a specific $H_0$ and $V$, was reported in \cite{km1, km2}).

The paper is organized as follows. Section II describes the response of the eigenvalues and eigenfunctions of a single chiral Hermitian matrix to an external perturbation preserving chiral symmetry. The response of an ensemble of such matrices, referred as  chiral Brownian ensembles, is considered in section III.
The information of sections II and III  is  used in section IV to derive the moments of the eigenvalues and eigenfunctions which in turn leads to an evolution equation for their joint probability density functions (JPDF); this is discussed in  section V. 
A measure specific integration of the evolution equations for the JPDF leads to  those for the spectral and strength fluctuation measures. In section VI and VII, we discuss only a few of them which have not been discussed before i.e the statistics of chiral eigenfunction and spectral correlations near chiral energy; a detailed analysis of other fluctuation measures will be presented elsewhere. In past, the spectral JPDF for non-chiral Hermitian ensembles has been obtained by mapping its diffusion equation to the Schrodinger equation of the Calogero Sutherland Hamiltonian (CSH). This motivates us to seek a similar mapping in the  present case and is discussed in section VIII.  This section also discusses the connection of chiral Brownian ensembles with three other important systems, namely, Wishart Brownian ensembles (WBE) \cite{pslg}, multiparametric Chiral ensembles \cite{psmulti} and bipartitie lattices \cite{ek} which appear in many areas of physics. The relevance of section VIII  lies in  the connections they reveal between different areas of physics in which ensembles of chiral and Wishart types as well as  CSH  are applicable.  We conclude in section IX with a brief summary of our results and open questions.

\section{Response of a Chiral Hermitian matrix to perturbation}

Prior to analysing their ensemble behavior, it is relevant to first discuss the properties of a single chiral matrix. 

\subsection{Eigenvalues and eigenfunctions of a chiral matrix}

The chiral symmetry  in a linear operator, if represented  in a basis preserving it,  manifest as the nonzero off-diagonal blocks
\begin{eqnarray}
H= \left( {\begin{array}{cc}   0  & C  \\  C^{\dagger} & 0 \end{array}} \right) .
\label{ch1}
\end{eqnarray}
where $C$ is a general a $N\times(N+\nu)$ complex matrix if $H$ has no other anti-unitary symmetry; as clear from above, $H_{k,N+l} = C_{kl}$. For cases with time-reversal  symmetry also present, $C$ is a real or quaternion matrix based on the presence/ absence of   
rotational symmetry (i.e integer or half integer angular momentum). The presence of additional symmetries and/or conservation laws can further impose constraints on the matrix elements which  affects the statistical behavior (\cite{pt1,psijmp}). For clarity purposes, here we confine our study only to $C$ real or complex with no other matrix constraints. The elements of  $C$ matrix  can then be written as $C_{kl}=\sum_{s=1}^{\beta} (i)^{s-1}C_{kl;s}$ where $k=1 \to N, l=1 \to (N+\nu)$ and $\beta=1$ or $2$ for $C$ real or complex; the subscript $s$ here refers to real ($s=1$) or imaginary part ($s=2$) of $C_{kl}$. (The generalization to quaternion $C$ can be done following similar steps but is technically tedious and is therefore not included here).

With $H$ given by eq.(\ref{ch1}), let $E$ be its eigenvalue matrix ($E_{mn} =e_n \delta_{mn}$) and  $U$ as the eigenvector matrix, with $U_{kn}$ as the $k^{th}$ component of the eigenvector $U_n$ corresponding to eigenvalue $e_n$. Following from eq.(\ref{ch1}), Tr($H$) is zero which then implies that the  eigenvalues  of $H$ exist  in equal and opposite pairs or are zero; let us refer such pairs as ${e}_n, e_{n+N}$ with $e_n=- e_{n+N}$, $1\le n \le N$.  Clearly the number of zero eigenvalues in $\nu$. Henceforth, the eigenvalues are labelled such that $e_k$, $k=1 \to N$ correspond to positive eigenvalues with their negative counterpart lying from $k=N+1 \to 2N$ and $k=2N+1 \to 2N+\nu$ refers to zero eigenvalues.

Writing the   eigenvector corresponding to $e_n$  as  $U_n=\left(\begin{array}{cc} X_n  \\  Z_n \end{array}\right)$ for $n=1 \to 2N+\nu$,  with $X_n, Z_n$ as column vectors with $N$ components and $N+\nu$ components respectively, the chirality implies $X_{N+k}=X_{k}, Z_{N+k}=-Z_k$ for $k=1 \to N$.  Eq.(\ref{ch1}) then gives $C \; Z_n = e_n \; X_n$ and $C^{\dagger} \;  X_n = e_n \; Z_n$ which leads to 
\begin{eqnarray}
C^{\dagger} C \; Z_n = e_n^2 \; Z_n, \qquad \quad C C^{\dagger} \; X_n = e_n^2 \; X_n  \qquad n=1 \to N
\label{wis1}
\end{eqnarray}
As clear from the above, $e_n$ are the singular values of matrix $C$ or $C^{\dagger}$. The orthogonality condition $U_n^{\dagger}.U_{n+N} =0$ along with normalization $U_n^{\dagger}.U_n=1$ further gives 
\begin{eqnarray}
X_n^{\dagger}. X_n ={1\over 2} ,  \qquad  Z_n^{\dagger}. Z_n ={1\over 2}   \qquad n=1 \to 2N.
\label{xx}
\end{eqnarray}

With  $U_{2N+k}$ as the eigenvector corresponding to zero eigenvalue $e_{2N+k} =0$ with $k=1 \to \nu$, one also has $C \; Z_{2N+k}= 0$, $C^{\dagger} C \; Z_{2N+k}=0$,  indicating  $Z_{2N+k}$ as the eigenvector of $C^{\dagger} C$ with zero eigenvalue; also note $Z_n^{\dagger} Z_{2N+k} = 0$ for $n=1 \to 2N$.

 Further one also has 
\begin{eqnarray}
C^{\dagger} \; X_{2N+k}=0,  \qquad CC^{\dagger} \; X_{2N+k}=0. 
\label{ze}
\end{eqnarray}
Note however that  $X_{2N+k}$  is not an eigenvector of $CC^{\dagger}$ and can be chosen as a null vector (without loss of generality as the choice satisfies both relations in eq.(\ref{ze})) which gives 
\begin{eqnarray}
U_{2N+k}=\left(\begin{array}{cc} 0  \\  Z_{2N+k} \end{array}\right), 
\qquad  Z_{2N+k}^{\dagger}. Z_{2N+k}=1
\label{uz}
\end{eqnarray}
Note the above choice also satisfies  the condition $U_n^{\dagger}.U_{2N+k} =0$.
Clearly the components of the state corresponding to zero energy, often referred as the  chiral state, are non-zero only in  one part of the basis.

\subsection{Variation of eigenvalues and eigenfunctions}

A random perturbation of $H$ affects, in general, the statistical behaviour of its eigenvalues and eigenfunctions. The effect can be determined by a prior knowledge of the response of $e_n$ and $U_n$ to change in an arbitrary  matrix element $H_{kl;s}$ (with subscript $s$ again referring to real ($s=1$) or imaginary part ($s=2$) of $H_{kl}$). Although the additional constraint of chirality either leads to zero eigenvalues or their equal and opposite pairs and various relations among their eigenfunctions (section II), the steps for the derivation for ${\partial e_n \over\partial H_{kl;s}}$ and ${\partial U_{rn} \over\partial H_{kl;s}}$ with $e_n, U_n$ as non-zero energy state remain essentially unaffected and the results can still be expressed in the same form as for a Hermitian matrix without chirality; the derivation for the latter case is discussed in \cite{pswf, ps-all}. To make the paper self-content, these steps are described in {appendix B} for the eigenvalue derivatives. As mentioned below, similar steps also lead to various derivatives of the eigenfunction components for $n < 2N$. The results for the zero energy i.e chiral state however can not be derived by the same route; the details are discussed in {\it appendix C}.

Differentiating the eigenvalue equation $H U=U E$ with respect to $H_{kl;s}$, subsequently applying orthogonality relations  for the  eigenfunctions  
$\sum_{k=1}^{2N+\nu} U_{km}^* U_{kn}= \delta_{nm}$
 $\sum_{k=1}^N U_{km}^* U_{kn}=(1/2) \delta_{nm}$,  $\sum_{l=1}^{N+\nu} U_{N+l, m}^* U_{N+l, n}=(1/2) \delta_{mn}$
 (see eq.(\ref{xx})),  with $U_{kn}^*= U_{kn}$ for case $\beta=1$, leads to 
\begin{eqnarray}
\frac{\partial e_n}{\partial H_{k,N+l;s}} &=& i^{s-1}[U_{kn}^* U_{N+l,n}+(-1)^{s+1} U_{N+l,n}^* U_{kn}] \label{ch9} \\
\frac{\partial U_{rn}}{\partial H_{k,N+l;s}} &=& \sum_{j=1}^{2N+\nu}  (T^{-1} )_{rj} \; X_j^{(n)}, 
\label{ch10}
\end{eqnarray}
where $T$ is a $(2N+\nu) \times (2N+\nu)$ matrix with elements 
\begin{eqnarray}
T_{rj} &=& \delta_{rj}-\sum_{m=2N+1}^{2N+\nu} U_{jm}^* U_{rm} 
= \sum_{m=1}^{2N} U_{jm}^* U_{rm} \qquad {\rm for} \; n > 2N \\
&=& \delta_{rj} \hspace{2.7in} {\rm for} \; n \le 2N
\label{ch11}
\end{eqnarray}
 and
\begin{eqnarray}
X_j^{(n)} = i^{s-1}\sum_{m=1, \atop m\neq n}^{2N+\alpha}\frac{U_{jm}}{e_n-e_m}[U_{km}^*U_{N+l,n}+(-1)^{s+1}U_{N+l,m}^*U_{kn}], 
\label{ch12}
\end{eqnarray}
with $\alpha=\nu$ if $n \le 2N$ and $\alpha=0$ if $n > 2 N$ (note $e_{2N+k}=0$ for $k=1 \to \nu$).

Further multiplication of eq.(\ref{ch9}) and eq.(\ref{ch10})  by $H_{kl;s}$, a summation over indices $k,l,s$ and  subsequent use of relations, $\sum_s i^{s-1} H_{kl;s}=H_{kl}$,  $\sum_s  (-i)^{s-1} H_{kl;s} =H_{kl}^* = H_{lk}$, $e_n \delta_{mn} = \sum_{i,j} U^*_{im} H_{ij} U_{jn}$ and interchange indices $k, l$ in one of the terms leads to 
\begin{eqnarray}
\sum_{k,l,s=1}^{N, N+\nu,\beta} \frac{\partial e_n}{\partial H_{k,N+l;s}}H_{k,N+l;s} &=& e_n       \label{ch13} \\
\sum_{k, l,s=1}^{N, N+\nu, \beta} \frac{\partial U_{pn}}{\partial H_{k,N+l;s}}H_{k,N+l;s} &=&  0
\label{ch14}
\end{eqnarray}
Here the superscripts on a summation sign follow the same order as in its subscripts. As example, the steps for eq.(\ref{ch13}) are illustrated 
in (see {\it appendix B}).

The relations involving second order derivatives of $e_n$ and $U_{rn}$ can similarly be derived.
Differentiating eq.(\ref{ch9}) and eq.(\ref{ch10}) again with respect to $H_{k,N+l;s}$ leads to second order ones in the left  and first order ones in the right side. 
Using eq.(\ref{ch9}) and eq.(\ref{ch10}) again, the latter can be rewritten in terms of eigenfunction components. This,  followed by a summation over indices $k,l,s$ and using relation $U^{\dagger} U=1$, results in  
\begin{eqnarray}
\sum_{k, l, s=1}^{N, N+\nu, \beta}  \frac{\partial^2 e_n}{\partial H^2_{k,N+l;s}}
&=& \beta\sum_{m=1 \atop m\neq n}^{2N+\nu} \frac{1}{e_n-e_m}  \hspace{1.0in} {\rm for} \; n \le 2N \label{ch15}\\
\sum_{k, l, s}^{N, N+\nu, \beta} \frac{\partial^2 U_{rn}}{\partial H^2_{k,N+l;s}} &=& -{\beta \over 2} \; \sum_{m=1 \atop \neq n}^{2N+\nu} \frac{U_{rn}}{(e_n-e_m)^2}
\hspace{1.0in} {\rm for} \; n \le 2N \label{ch16}
\end{eqnarray}
The derivation for eq.(\ref{ch15}) is discussed in appendix B; eq.(\ref{ch16}) for $U_n$ with $n \le 2N$ can be derived by similar steps. The case for $U_{2N+k}$, $k=1 \to \nu$ however requires different derivation and, as  discussed in {\it appendix} C, can be given as
\begin{eqnarray}
\sum_{k, l, s}^{N, N+\nu, \beta} \frac{\partial^2 U_{rn}}{\partial H^2_{k,N+l;s}} = -{\beta \over 2} \; \sum_{m=1 \atop \neq n}^{2N} \frac{U_{rn}}{(e_n-e_m)^2}  \hspace{1.0in} {\rm for} \; n > 2N
\label{chh16}
\end{eqnarray}

A product of eq.(\ref{ch9}) for two different energies, say $e_n, e_m$ or its product with eq.(\ref{ch10}) gives a sum over terms containing products of eigenfunction components. A  summation over indices $k,l,s$ and using relation $U^{\dagger} U=1$ then leads to (see {\it appendix B})
\begin{eqnarray}
\sum_{k,l,s=1}^{N, N+\nu, \beta}\frac{\partial e_n}{\partial H_{k,N+l;s}}\frac{\partial e_m}{\partial H_{k,N+l;s}} &=& 2 \; \delta_{mn} \quad {1 \le m,n \le N} \label{ch12a}\\
\sum_{k, l, s=1}^{N, N+\nu, \beta}\frac{\partial e_n}{\partial H_{k,N+l;s}}\frac{\partial U_{pn}}{\partial H_{k,N+l;s}}=0
\label{ch17}
\end{eqnarray}
Similarly a product of eq.(\ref{ch10}) for two  eigenfunction components, say $U_{rn}, U_{pt}$ for $n, t \le 2n$ and, proceeding as in the previous case gives (see {\it appendix C} for $n,t >2N$)
\begin{eqnarray}
\sum_{k, l, s}^{N, N+\nu, \beta}  \frac{\partial U_{pt}}{\partial H_{k,N+l;s}}\frac{\partial U_{rn}}{\partial H_{k,N+l;s}} &=& -{ \beta \over 2}  \; \frac{S_{pnrt}}{(e_n-e_t)^2}(1-\delta_{nt}) 
\label{ch18a}\\
\sum_{k, l, s=1}^{N,N+\nu, \beta} \sum_{s=1}^{\beta} \frac{\partial U_{pt}}{\partial H_{k,N+l;s}}\frac{\partial U^*_{rn}}{\partial H_{k,N+l;s}} &=& {  \beta \over 2} \; \sum_{m=1 \atop m \not=r}^{2N+\alpha} \frac{D_{pmrm}}{(e_m-e_n)^2} \; \delta_{nt}
\label{chh18}
\end{eqnarray}
where 
\begin{eqnarray}
S_{pnrt} &=& U_{pn} \; (T^{-1} U)_{rt}  \qquad   {\rm for} \;\;  t \le 2N, n > 2N, \nonumber \\
&=&  (T^{-1} U)_{pn} \; U_{rt}  \qquad   {\rm for} \;\;  t > 2N, n \le 2N, \nonumber \\
                &=& U_{pn} U_{rt}  \qquad  \qquad  {\rm for} \;\;   t, n \le 2N,  
\label{ss}              
 \end{eqnarray}              
    and 
\begin{eqnarray}
D_{pmrm} &=& (T^{-1} U)_{pm} \;  (T^{-1} U)^*_{rm}   \qquad  {\rm for} \;\;  t, n > 2N \nonumber \\
 &=&  U_{pm} U^*_{rm}  \qquad  \qquad  {\rm for}\; \;  t, n \le 2N.
 \label{dd}
 \end{eqnarray}           
 Further    $S_{pnrt}=0$ if both $n,t >2N$ and $D_{pmrm}=0$ if $r \le 2N, t > 2N$ or $r > 2N, t \le 2N$.

It is worth emphasizing here that eqs.(\ref{ch9}-\ref{chh18}) are exact and form the backbone of this paper. Further, for a compact presentation,  eqs.(\ref{ch8}-\ref{chh18})  are written in a form similar to their non-chiral counterparts (discussed in {\it appendix B} of \cite{pswf}) with chirality constraint assumed to be implicit.  The difference between the two cases becomes apparent when  chirality constraint is explicitly imposed. (For example,  eq.(\ref{ch16}) now has two parts: $\sum_{k, l, s}^{N, N+\nu, \beta} \frac{\partial^2 X_{rn}}{\partial H^2_{k,N+l;s}} = -{\beta \over 2} \; X_{rn} \; \Phi_n $, $r=1 \to N$ and $\sum_{k, l, s}^{N, N+\nu, \beta} \frac{\partial^2 Z_{pn}}{\partial H^2_{k,N+l;s}} = (-1)^q \; {\beta \over 2} \; Z_{pn} \; \Phi_n$ for $p= 1 \to N+\nu$, $q=0$ for $n <N$ and $q=1$ for $n >N$ and $\Phi_n = \sum_{m=1}^{N} \frac{e_n^2+e_m^2}{(e_n^2-e_m^2)^2} + 2 \sum_{m=1}^{N} {1\over e_m^2} $ ).

\section{Ensemble density and its diffusion}

Consider an arbitrary $M \times M$  chiral matrix $H_0$ subjected to a random perturbation, of strength $t$, by another $M \times M$   chiral matrix $V$ with $M=2N+\nu$:
\begin{eqnarray}
H_0= \left( {\begin{array}{cc}   0  & C_0  \\  C_0^{\dagger} & 0 \end{array}} \right), \qquad   V=\left( {\begin{array}{cc}   0  & C_v  \\  C_v^{\dagger} & 0 \end{array}} \right) .
\label{ch2}
\end{eqnarray}

The perturbed matrix $H(t)$ is described  as $H(t)=\sqrt{f} (H_0+t \; V)$ with $f=(1+ \gamma t^2)^{-1}$, $H(0)=H_0$ as a fixed random matrix and $\gamma$ as an arbitrary positive constant.  
Using eq.(\ref{ch2}), the distribution, say $\rho(H)$, of the elements of the matrix $H$ can be expressed in terms of those of $C$:
\begin{eqnarray}
\rho(H) = \rho_c(H_{k,N+l}) \; F_c \; F_h
\label{rhoh}
\end{eqnarray}
with $F_h(H)=\delta(H-H^{\dagger})$ and $F_c=\left(\prod_{k,l=1}^N \; \delta(H_{kl}) \right)$ as the constraints due to Hermiticity and chirality of $H$, respectively, and $\rho_c(C)$ as the probability density of the ensemble of $C$ matrices.

Assuming the matrix elements of $C_0$ and $C_v$  distributed with the probability densities $\rho_0(C_0)$ and $\rho_v(C_v)$, the probability density $\rho_c(C) =\langle \delta\left(C-\sqrt{f} (C_0+t \; C_v) \right) \rangle $ of the $C$-ensemble is given by (with $\langle \rangle$ as the ensemble average)
\begin{eqnarray}
\rho_c(C) 
&=& \int \; \rho(C, t| C_0,0) \; \rho_0(C_0) \; {\rm D}C_0
\label{rhoc}
\end{eqnarray}
with  
\begin{eqnarray}
\rho(C, t|C_0,0) &=& \int \;  \delta\left(C-\sqrt{f} (C_0+t \; C_v) \right)\; \rho_v(C_v) \; {\rm D}C_v\nonumber \\
&=&
\left(\frac{1}{t \sqrt{f}}\right)^{N(N+\nu)} \; \rho_v \left(\frac{C-\sqrt{f} C_0}{t \sqrt{f}} \right);
\label{rhoc1}
\end{eqnarray}  
here $C=C_0$ for $t \rightarrow 0$, $C \rightarrow {C_v\over \sqrt{\gamma}}$ for $t \rightarrow \infty$.  Following from above, a variation of $t$ then leads to a diffusion of $\rho_c(C)$ .  The  dynamics retains its Markovian character  if considered in terms of a rescaled evolution parameter $Y=-{1\over 2 \gamma} \;  \ln f ={1\over 2 \gamma} \; \ln (1+ \gamma \; t^2)$ \cite{sp}:
\begin{eqnarray}
C(Y)  &\equiv&  C(0) \; {\rm e}^{-\gamma Y} +  C_v \; \left({1- {\rm e}^{-2 \gamma Y} \over \gamma}\right)^{1/2}
 \label{att0} 
\end{eqnarray}

Assuming $V$ taken from Ch-GOE or Ch-GUE, the ensemble density $\rho_v(C_v) $ can be expressed as 
\begin{eqnarray}
\rho_v(C_v)  &=&  \mathcal{N} \;  {\rm exp}\left[-  \;{1\over 2 v^2} \; {\rm Tr} \left( C_v^a.C_v \right) \right]
\label{rhov}
\end{eqnarray}
where $C^a \equiv C^T$ or $C^{\dagger}$ .
A substitution of eq.(\ref{rhov}) in eq.(\ref{rhoc1}) leads to  $\rho(C, Y| C_0,0)$ as a Gaussian. This can also be seen as follows: a convolution of two Gaussians being another Gaussian, one can write, for a small increment of perturbation strength at $Y$ with $\rho_v(C_v)$ given by eq.(\ref{rhov}),  
\begin{eqnarray}
C(Y+\delta Y)  &\equiv&  {C(Y) + \sqrt{2 \; \delta Y} \; C_v(Y) \over \sqrt{1 + 2 \; \gamma \; \delta Y} } \label{att} \\
&\approx &  C(Y) \; \left(1 -  \gamma \; \delta Y \right) +  \sqrt{2  \; \delta Y} \; C_v(Y) + O((\delta Y)^{3/2}). 
\label{atm1}
\end{eqnarray} 
with symbol $''\equiv''$ implying the  equivalence not only of the  matrices on two sides but  their ensembles too.   (The equivalence of  eq.(\ref{att0}) and eq.(\ref{att}) along with the derivation of the diffusion equation for $C(Y)$ is discussed in \cite{sp}. )

An ensemble average (denoted by symbol $\langle. \rangle$) of eq.(\ref{atm1}) gives
\begin{eqnarray}
\langle{\delta C_{kl;s}} \rangle &=& -   \gamma \; C_{kl;s}  \; \delta Y, \label{amn2}  \\
\langle{\delta C_{kl;s} \; \delta C_{mn;s'}}\rangle
&=&  2  \; v^2 \; \delta_{km} \; \delta_{ln} \; \delta_{ss'} \; \delta Y
\label{amn3}
\end{eqnarray}
A substitution of the above moments in the  standard Fokker-Planck equation then leads to
\begin{eqnarray}
{1\over v^2}  \frac{\partial\rho_c}{\partial Y} =  \sum_{k,l,s} \frac{\partial}{\partial C_{kl;s}} 
\left[ \frac{\partial }{\partial C_{kl;s}}+{\gamma\over v^2} \; C_{kl;s}  \right]\; \rho_c
\label{ch8}
\end{eqnarray}

Eq.(\ref{rhoh}) along with the substitution of $C_{kl;s} =H_{k N+l;s}$ in the above then leads to the $Y$-governed evolution equation for $\rho(H)$.
\begin{eqnarray}
{1\over v^2} \frac{\partial\rho}{\partial Y} =  \sum_{k,l,s} \frac{\partial}{\partial H_{k N+l;s}}\left[\frac{\partial }{\partial H_{k N+l;s}}+{\gamma\over v^2} \; H_{k N+l;s} \right]  \rho.
\label{chi8}
\end{eqnarray}
The dynamics approaches to equilibrium as $Y \to \infty$ or $\frac{\partial\rho}{\partial Y} \to 0$; the solution of eq.(\ref{chi8}) in this limit corresponds to the ensemble density for ch-GUE or ch-GUE. 

Eq.(\ref{chi8}) describes the evolution of the ensemble density  $\rho(H)$ under a chirality preserving perturbation from an arbitrary initial condition $\rho_0(H_0)$  to final state as ch-GOE or ch-GUE, for $C$  as a rectangular real or complex matrix. A special case of the above evolution  for a specific initial condition i.e GUE and with $C$ as a Hermitian matrix  is analysed in \cite{km1}.


Eq.(\ref{chi8}) can further be used to derive the $Y$ governed evolution of the probability densities of the eigenvalues and eigenfunctions of $H$; this is discussed in next section.

\section{Moments of eigenvalues and eigenfunctions}

 A small change $\delta Y$ in $Y$ subjects $H$ and thereby its non-zero eigenvalues $e_n$ and corresponding eigenfunctions $U_n$ to undergo a diffusion. 
 As the dynamics preserves chirality, the zero eigenvalues of $H$ remain unaffected but corresponding eigenfunctions $U_{2N+k}$ change with changing $H$ to ensure $H U_{2N+k}=0$ with $k=1 \to \nu$. Assuming Markovian dynamics, the diffusion equation for the eigenvalues and eigenfunctions can be derived if their moments are known. 
The standard Fokker-Planck approach  on which it is based can in general be described as follows. 
Assuming Markovian process, the parametric diffusion of the  joint probability distribution $P_x(x_1,\ldots, x_N; Y)$ of $N$ variables $x_n$, $n =1,\ldots, N$ from an arbitrary initial condition, with $Y$ as the parameter, is given by 
\begin{eqnarray}
{\partial P_x\over\partial Y} \; \delta Y = {1\over 2} \sum_{k,l=1}^N {\partial^2 \over \partial x_{k} \partial x_l} \; (\langle\delta x_k \delta x_l \rangle\; P_x)  -\sum_{k=1}^N  {\partial \over \partial x_k} \; (\langle \delta x_k \rangle \; P_x)
\label{px}
\end{eqnarray}
Replacements $x_n \to H_{k,N+l;s}$ and $P_x \to \rho(H)$, followed by a comparison of the above equation with eq.(\ref{chi8}) leads to the moments of $H$-matrix elements
\begin{eqnarray}
\langle{\delta H_{k,N+l;s}} \rangle &=& -   \gamma  \; H_{k,N+l;s}  \; \delta Y, \label{amn2a}  \\
\langle{\delta H_{k,N+l;s} \; \delta H_{m,N+n;s'}}\rangle
&=&  2 \; v^2 \; \delta_{km} \; \delta_{ln} \; \delta_{ss'} \; \delta Y
\label{amn3a}
\end{eqnarray}
with $\langle . \rangle$ implying an ensemble average.

Our next step is to derive the moments for the eigenvalues and eigenfunctions from the above. 
A standard route in this context  is 2nd order perturbation theory of Hermitian matrices but to avoid the issues with treatment of degenerate chiral  eigenfunctions, we pursue an alternate route, described as follows.

The change in an arbitrary function $f(H)$ of the matrix elements of $H$, due to latter's variation, can be   described by the Taylor's series 
\begin{eqnarray}
\delta f= \sum_{k,l;s}   \frac{\partial f}{\partial H_{kl;s}} \delta H_{kl;s} 
+{1\over 2}  \sum_{k,l;s}\sum_{a,b;s'}   \frac{\partial^2 f}{\partial H_{kl;s} \partial H_{ab;s'}} {\delta H_{kl;s} \delta H_{ab;s'} }  + O((\delta H_{kl;s})^3)
\label{fa1}
\end{eqnarray}
An ensemble average of the above  can be written as 
\begin{eqnarray}
\langle \delta f \rangle = \sum_{k,l;s}   \frac{\partial f}{\partial H_{kl;s}} \; \langle \delta H_{kl;s} \rangle\; 
+ {1\over 2} \sum_{k,l;s}\sum_{a,b;s'}   \frac{\partial^2 f}{\partial H_{kl;s} \partial H_{ab;s'}} \langle{\delta H_{kl;s} \delta H_{ab;s'} }\rangle  + O((\delta H_{kl;s})^3)
\label{fa2}
\end{eqnarray}
Substitution of eq.(\ref{amn2a},\ref{amn3a}) in the right side of the above equation then leads to
\begin{eqnarray}
\langle \delta f \rangle = -\gamma \; \sum_{k,l;s}   \frac{\partial f}{\partial H_{kl;s}} \; H_{kl;s} \; \delta Y 
+  \sum_{k,l;s} \frac{\partial^2 f}{\partial H_{kl;s}^2} \; \delta Y  + O((\delta H_{kl;s})^3)
\label{fa3}
\end{eqnarray}
Further keeping terms only upto first order of $\delta Y$, one can similarly write, for the  function $f(H)$ and $g(H)$,  
\begin{eqnarray}
\langle \delta f \; \delta g \rangle = \sum_{k,l;s}   \frac{\partial f}{\partial H_{kl;s}} \;\frac{\partial g}{\partial H_{kl;s}} \; \delta Y 
\label{fa4}
\end{eqnarray}

Eqs.(\ref{fa3}), (\ref{fa4}) are applicable in general for any arbitrary function of $H$ to eqs.(41)-(45). Thus choosing $f$ and $g$ as the eigenvalues and/or the eigenfunction components  of $H$, one can derive the $1^{st}$ and $2^{nd}$ order moments of their various combinations. For example,  replacing $f=e_n$ and $g=e_m$ in eqs.(\ref{fa3}, \ref{fa4}) and subsequently using eq.(\ref{ch13}) and eq.(\ref{ch15}), we have

 \begin{eqnarray}
\langle{\delta e_n} \rangle
&=& \beta \; v^2 \; \left[ - {\gamma \over \beta v^2} \; e_n + \frac{\nu+1/2}{ e_n} + \sum_{m=1,m\not=n}^{N} 
\; {2 e_n\over e_n^2-e_m^2}\right] \delta Y \nonumber \\
\langle{\delta e_n \delta e_m }\rangle &=& 
2 \; v^2 \; \delta_{nm} \; \delta Y 
\label{enm}
\end{eqnarray}
 As the perturbation is assumed to preserve the chiral symmetry leaving zero eigenvalues unchanged, this implies: $\langle{\delta e_{2N+k}} \rangle =0$  for $k=1 \to \nu$.
 
The moments of eigenfunctions can similarly be derived by replacing $f=U_{pt}$ and $g=U_{rn}$ or $U^*_{rn}$ in eqs.(\ref{fa3},\ref{fa4}) and subsequently using eq.(\ref{ch14})-eq.(\ref{chh18}), 
\begin{eqnarray}
\langle{\delta U_{jn}}\rangle &=&  - \beta \; v^2 \; \alpha_1^{(n)} \;  \sum_{m=1,\atop m\not=n}^{2N+ \nu} {  U_{jm}  \over (e_n-e_m)^2}  \; \delta Y 
\label{u1}
\end{eqnarray}
with $\alpha_1^{(n)}=0$ if $n > 2N$ and $\alpha_1^{(n)}=1$ if $n \le 2N$ and
\begin{eqnarray}
\langle{\delta U_{pt} \; \delta U_{rn} } \rangle &=& - \beta \; v^2 \;  {S_{pnrt} \over (e_t-e_n)^2}  \; (1-\delta_{nt}) \;  \delta Y 
\label{u3}
\end{eqnarray}
and 
\begin{eqnarray}
\langle{\delta U_{pt} \; \delta U^*_{rn} } \rangle &=&  \beta \; v^2  \;  \sum_{m=1,\atop m\not=n}^{2N+\alpha} {D_{pmrm}  \over (e_n-e_m)^2}  \;  \delta_{nt} \;  \delta Y  
\label{u4}
\end{eqnarray}
with  $S_{pnrt}, D_{pmrm}$ given by eq.(\ref{ss}), eq.(\ref{dd}) and angular brackets implying conditional ensemble averages with fixed $e_j, U_j$, $j=1,\ldots, 2N+\nu$. 
 Further, to first order in $\delta Y$,  the ensemble averaged correlation between $\delta e_k$ and $\delta U_{jn}$ is zero (for both $\beta=1$ or $2$): 
\begin{eqnarray}
\langle{\delta e_k \; \delta U_{jn}}\rangle =  0 \qquad \forall k,j,n
\label{f7}
\end{eqnarray}

\section{Diffusion of spectral and strength JPDF}

With $\{U_n\}$ and $\{e_n\}$ referring to the sets of all eigenvectors
and eigenvalues for $n=1 \to 2N+\nu$, an appropriate definition 
of their JPDF for a chiral Hermitian ensemble can be given as 
\begin{eqnarray}
P^{(c)}_{ef,ev} (\{U_n\}, \{e_n \}) =P_{ef,ev} (\{U_n\}, \{e_n \}) \; F_{ef} \; F_{ev}
\label{pcf}
\end{eqnarray}
 with $F_{ef}$ and $F_{ev}$ as the chiral-Hermitian conditions on the eigenfunction components and eigenvalues,
\begin{eqnarray}
F_{ef} &\equiv& \delta(U^{\dagger} U-1) \; \prod_{k,n=1}^N \delta(U_{kn}-U_{k,N+n}) \prod_{l,n=1}^N \delta(U_{N+l,n}+U_{N+l,N+n}) \prod_{s,m=1}^{N,\nu} \delta(U_{s,2N+m}), \nonumber \\
F_{ev} &=&\prod_{n=1}^N \delta(e_n + e_{N+n}) \prod_{k=1}^{\nu} \delta(e_{2N+k})
\label{fef}
\end{eqnarray}

Our next step is to derive the $Y$-governed diffusion equation for the  density $P_{ef,ev} (\{U_n\}, \{e_n \})$. Following standard Fokker-Planck approach described by eq.(\ref{px}), the diffusion can be described as follows
%

\begin{eqnarray}
{1\over v^2} {\partial P_{ef,ev} \over\partial Y}  &=& (  {\mathcal L}_U +  {\mathcal L}_U^*+  {\mathcal L}_E) P_{ef,ev} 
\label{f8} 
\end{eqnarray}
where $ {\mathcal L}_U$ and $ {\mathcal L}_E$ refer to two parts of the Fokker-Planck operator corresponding to eigenvalues and eigenfunction components, respectively,

\begin{eqnarray}
 {\mathcal L}_U \; \delta Y &=& {\beta \over 2}\sum_{j,n=1}^{2N+\nu} {\partial \over \partial U_{jn}}
\left[{\beta\over 4} \sum_{k,l=1}^{2N+\nu} \left({\partial \over \partial U_{kl}} \langle \delta U_{jn} \delta U_{kl} \rangle
+ {\partial \over \partial U^*_{kl}} \langle \delta U_{jn} \delta U^*_{kl} \rangle\right)
- \langle \delta U_{jn} \rangle \right] 
\label{f9}
\end{eqnarray}
and $ {\mathcal L}_E$ is
\begin{eqnarray}
 {\mathcal L}_E \; \delta Y &=& \sum_{n=1}^{2N+\nu} {\partial \over \partial e_n}
\left[{1\over 2} \; {\partial \over \partial e_n} \langle (\delta e_n)^2 \rangle - \langle \delta e_n \rangle \right] 
\label{f9p}
\end{eqnarray}
with $P_{ef,ev}$  subjected to following boundary condition: 
$P_{ef,ev} \to 0$ for $|U_{jn} |\to \pm 1/\sqrt{2}, e_n \to (-\infty, \infty )$ for $n=1 \to 2 N$ and $|U_{j,2N+k} |\to \pm 1$ for $k=1 \to \nu$ with $j=1 \to 2N+\nu$; these conditions on the eigenfunction components  follow from the orthogonality relations eq.(\ref{ze}) and eq.(\ref{uz}). A substitution of the moments (eqs.(\ref{enm}, \ref{u1}, \ref{u3}, \ref{u4})) in eq.(\ref{f9}) and eq.(\ref{f9p}) then leads to  
\begin{eqnarray}
& &{\mathcal L}_{U} =  \sum_{k,l,j,n; n \not=l}  {\partial^2 \over \partial U_{jn} \partial U_{kl}}
\left( { S_{jlkn} \over (e_l-e_n)^2} \right) + \sum_{m=1,m\not=n}^{2N+\alpha} \sum_{k,j,n} \; {\partial^2 \over \partial U_{jn} \partial U_{kn}^*} \left ( {D_{jmkm} \over (e_n-e_m)^2} \right)  \nonumber \\
&+&   \sum_{m=1,m\not=n}^{2 N+\nu} \sum_{j,n=1}^{2N}  {\partial \over \partial U_{jn} }  \left({ U_{jn} \over (e_n-e_m)^2} \right)
\label{defev}
\end{eqnarray}
and 
\begin{eqnarray}
{\mathcal L}_{E} = \sum_{n=1}^{2N+\nu} \frac{\partial}{\partial e_n}\left[\frac{\partial}{\partial e_n}- \sum_{m=1}^N \frac{\beta }{e_n- e_m} + {\gamma\over v^2} \; e_n\right].
\label{dev}
\end{eqnarray}
As mentioned in section III, here $\gamma$ is an arbitrary positive constant.

It is important to note here that eq.(\ref{f8}) (along with eqs.(\ref{f9}, \ref{f9p})) does not take into account the symmetry constraints (i.e chirality and Hermiticity for our case). Eqs.(\ref{defev}, \ref{dev}) are also obtained by a substitution of the  moments given by eqs.(\ref{enm}-\ref{f7}). As the latter are implicitly assumed to be subjected to the constraints, so is the solution of eq.(\ref{f8}). To avoid technical issues  at a later stage (e.g. while calculating fluctuations) however, $P_{ef,ev}$ should be multiplied with chiral and Hermitian constraints explicitly as in eq.(\ref{pcf}). 

As discussed next, an integration  of eq.(\ref{f8}) over all undesired variables  will now lead to an evolution equation for the JPDF of the desired combination  of  eigenfunctions and eigenvalues.

\subsection{Strength JPDF}

Let us define  $P_{ef}(\{U_{rn} \}; Y)$ as the JPDF of all components of the  eigenvectors $U_1, U_2, \ldots,  U_{2N+\nu}$. It s related to $P^{(c)}_{ef,ev}$:
\begin{eqnarray}
P_{ef}(\{U_{rn} \}; Y)= P_{ef}(U_1, \ldots, U_{2N+\nu}) =\int P^{(c)}_{ef,ev}(\{e_n, U_n \}) \; \prod_{k=1}^{2N+\nu} {\rm d}e_n.
\label{peef1}
\end{eqnarray}

As explained in {\it appendix D}, an integration over all eigenvalues of eq.(\ref{f8}) along with above definition then leads to diffusion equation for  $P_{ef}$ 
\begin{eqnarray}
{1\over v^2} \frac{\partial P_{ef}}{\partial Y} &=& {\beta \over 2}\left[{\mathcal L}_{I} + {\mathcal L}_{I}^*\right] P^{(c)}_{ef,ev}
\label{chh19}
\end{eqnarray}
with 
\begin{eqnarray}
& &{\mathcal L}_{I} P^{(c)}_{ef,ev} =  \sum_{k,l,j,n; n \not=l}  {\partial^2  I_{jlkn}\over \partial U_{jn} \partial U_{kl}}
+ \sum_{k,j,n} \; {\partial^2 J_{jkn}\over \partial U_{jn} \partial U_{kn}^*}
 +   \sum_{j,n=1}^{2N}  {\partial K_{jn} \over \partial U_{jn} }  
\label{chx20}
\end{eqnarray}
with $I_{jlkn} =\int { S_{jlkn} \; P^{(c)}_{ef,ev} \over (e_l-e_n)^2}  {\rm D}e$, 
$J_{jkn} = \sum_{m=1,m\not=n}^{2N+\alpha}  \int   { \; D_{jmkm} \; P^{(c)}_{ef,ev} \over (e_n-e_m)^2} \; {\rm D}e,
K_{jn} = \sum_{m=1,m\not=n}^{2 N+\nu} \int { U_{jn} \; P^{(c)}_{ef,ev}\over (e_n-e_m)^2} \; {\rm D}e$ with
${\rm D}e \equiv \prod_{n=1}^{N} {\rm d}e_n$. (Note as ${\mathcal L}_{I} = {\mathcal L}_{I} ^*$ for $\beta=1$, the factor ${\beta \over 2}$ in right side of eq.(\ref{chh19}) is needed  to keep  the correct number of the terms).

Here again in the stationary limits  of eq.(\ref{chh19}) i.e $ {\partial P_{ef} \over  \partial Y}$, $P_{ef}$ is expected to approach the JPDF of eigenfunctions for stationary chiral ensemble: $P_{ef}=F_{ef}$ (latter given by eq.(\ref{fef})).

\subsection{Spectral JPDF}

Let us define  $P_{ev}(\{e_n \}; Y)$ as the JPDF of all eigenvalues $e_1, e_2, \ldots,  e_{2N+\nu}$;  
it can be derived from $P^{(c)}_{ef,ev}$ as follows.
\begin{eqnarray}
P_{ev}(\{e_n \}; Y)= \int P^{(c)}_{ef,ev}(\{e_n, U_n \}) \; \prod_{k=1}^{2N+\nu} {\rm d}U_n.
\label{peev1}
\end{eqnarray} 
For technical purposes, it is easier to express the above in terms of the JPDF $P_N(e_1, e_2, \ldots,  e_N)$ of the positive eigenvalues only, 
\begin{eqnarray}
P_{ev}(e_1, e_2, \ldots,  e_{2N+\nu})  \equiv P_N(e_1, e_2, \ldots,  e_N) \; \prod_{k=1}^N  \delta(e_k + e_{k+N}) \; \prod_{n=1}^{\nu} \delta(e_{2N+n}).
\label{pe1}
\end{eqnarray}
 
 An integration of eq.(\ref{f8}) over all eigenfunctions  leads to the diffusion equation for  $P_N$ ({\it appendix E})
\begin{eqnarray}
{1\over  v^2} \; \frac{\partial P_{N}}{\partial Y}=  2 \sum_{n=1}^N \frac{\partial}{\partial e_n}\left[\frac{\partial}{\partial e_n}- \frac{\beta \nu_0}{ e_n} - \sum_{m=1}^N \frac{2 \beta e_n}{e_n^2- e_m^2} + {\gamma\over v^2} \; e_n\right]P_{N}
\label{ch19}
\end{eqnarray}
with $\nu_0=\nu+1/2$. Here the pre-factor $2$ in right side results from the same contribution of the negative eigenvalues.

As mentioned in section II, $\lim_{Y \to \infty} \rho(H, Y)$ corresponds to the stationary chiral ensembles. For the stationary limits  of eqs.(\ref{ch19}) i.e ${\partial P_{ev} \over  \partial Y}$, $P_{ev}$  is therefore expected to approach corresponding JPDF of eigenvalues.
 It is easy to verify that, with $P_{ev}$ given by eq.(\ref{pe1}),  the stationary limit gives 
\begin{eqnarray}
P_N(e_1, e_2, \ldots,  e_N) &=& {\mathcal N} \; \prod_{n=1}^N e_n^{\beta(\nu+1/2)}  \; \prod_{j > k}^N |e_j^2 - e_k^2|^{\beta} \; {\rm e}^{-{\gamma \over 2\; v^2} \sum_{j=1}^N e^2_j }.
\label{stdis1}
\end{eqnarray}
with ${\mathcal N}$ as the normalization constant.

A general solution of eq.(\ref{ch19}) for arbitrary initial condition is desirable.   As discussed latter in section VIII.A, it can be obtained, in principle, by the mapping to a general state of a Calogero-Sutherland Hamiltonian. 
The details are  technically complicated and beyond the scope of present analysis. An alternative route, discussed in \cite{km1}  for a specific initial condition i.e GUE and with $C$ as a Hermitian matrix (thus implying $\nu=0$), is also in principle applicable for other initial conditions but seems equally complicated. Note however the relevance of eq.(\ref{ch19}) lies beyond its solution; as mentioned in section VIII.C, it reveals the connection of chiral Brownian ensemble with other ensembles.

\section{Eigenfunction Fluctuation Measures}

Eq.(\ref{chh19}) describes the $Y$-governed evolution of the JPDF of all $2N+\nu$ eigenfunctions of $H$. Its integration over  a few of the eigenfunctions leads to $Y$-dependent diffusion equations for the  JPDfs  of the rest of them.  Various fluctuations measures for these eigenfunctions (referred also as strength measures) e.g. local intensity distribution or inverse participation ration can then be derived from these equations (see e.g. \cite{pslg}). A detailed calculations for many measures can however be avoided by noting the analogy of their diffusion equations with already known cases.  For example, writing $U_n$ in terms of$X_n, Z_n$ (see section II.A), we have 
\begin{eqnarray}
 P_{X}(X_1,.., X_{N})  = \int P_{ef}(U_1, \ldots, U_{2N+\nu}) \; {\rm D} Z 
\label{pef1}
\end{eqnarray}
with ${\rm D}Z$ referring to an integration over all components of the eigenfunctions $Z_k$, $k=1 \to 2N+\nu$. Differentiating the above equation with respect to $Y$ and using eqs.(\ref{chh19}, \ref{chx20}), one can derive the diffusion equation for $X$-components only. 
But  as discussed latter in section VIII B (see eq.(\ref{wis1})), $X_n$ of $U_n$  is analogous to the eigenfunction of  Wishart matrix $L_1=C.C^{\dagger}$. The same analogy is then applicable between the diffusion equations for $P_X$  and  the JPDF of the eigenfunctions of Wishart Brownian ensemble (with same $Y$ for both cases).  Consequently the fluctuation measures derived in \cite{pslg} can directly be used for $P_x$ of the chiral ensemble discussed here.  Following the same logic, the fluctuation measures of the $Z$-components of the eigenfunctions of  $H$-matrix are same as those of $L_2=.C^{\dagger}.C$. Although these fluctuations for $L_2$ are not explicitly discussed in past, the steps are essentially the same as for $L_1$ in \cite{pslg}.  As an example, here we discuss, in detail, the statistical behavior of chiral eigenfunction $U_{2N+k}$, $k=1 \to \nu$.

\subsection{Diffusion of Chiral States}

 The eigenfunctions corresponding to zero eigenvalues, say $e_{2N+n}=0$ for $n=1 \to \nu$ are  often referred as zero modes or chiral states. As reported by many studies e.g. \cite{been-m, smb}, the statistics  of their components  is an important fluctuation measure.

 Consider the JPDF $P_{\nu}(U_{ 2N+1}, \ldots, U_{2N+\nu}) \equiv P_{\nu}(\{U_{k,2N+1} \}, \ldots, \{U_{k, 2N+\nu}\})$ of all  non-zero components of the eigenfunctions $U_{2N+n}$, $n=1 \to \nu$ given by eq.(\ref{uz}) (with $P_{\nu}$ referring to the JPDF of $\nu$ eigenfunctions and notation $\{x_k \}$ refers to the set of $N+\nu$ variables $ x_1,\ldots, x_{N+\nu}$).

Similar to derivation of eq.(\ref{chh19}) from eq.(\ref{f8}), the diffusion equation for $P_{\nu}$ can be obtained from eq.(\ref{chh19}) by its integration  over all other eigenvectors except those corresponding to zero eigenvalue. Noting that 
(i) $P_{\nu}$ and its derivatives become zero at the integration limits $|U_{rn}| \to \pm 1 \; \forall \; r, n=1 \to 2N+\nu$, 
(ii) $S_{jnkl}=0$ for both $n,l > 2N$, (iii)  on integration,  the terms with $S_{jnkl}$ with $n$ and/or $l \le 2N$ do not contribute, 
(iv) with the perturbation preserving chiral symmetry, the components $U_{k 2N+\eta}=0$ for $k=1 \to N$ during diffusion, for $\eta =1 \to \nu$ (see eq.(\ref{uz})), the integration leads to

\begin{eqnarray}
{1\over v^2} \; {\partial P_{\nu} \over \partial Y} = 
 {\beta \over 2} \; \sum_{\eta=2N+1}^{2N+\nu} \; \sum_{k,j=N+1}^{2N+\nu} \left({\partial^2 Q_{jk} \over \partial U_{j\eta} \partial U_{k \eta}^*} +{\partial^2 Q_{jk}^*\over \partial U_{j\eta}^* \partial U_{k\eta}} \right) 
\label{uzp}
\end{eqnarray}
where 
\begin{eqnarray}
Q_{jk} &=&  \sum_{m=1}^{2N} \int {U_{jm} \; U^*_{km}  \over e_m^2} \; P_{ef} \; \prod_{n=1}^{2N} \left({\rm d}U_n \; {\rm d}e_n\right)\nonumber \\
&=&  \sum_{m=1}^{2N} \int {U_{jm} \; U^*_{km}  \over e_m^2} \; P_{\nu+1} \; {\rm d}U_m \; {\rm d}e_m 
\label{qjk1}
\end{eqnarray}
with $P_{\nu+1} \equiv P_{\nu+1}(e_m, U_m, U_{2N+1}, \ldots, U_{2N+\nu})$.

Assuming weak correlations between the chiral and non-chiral states (as the former are not affected by perturbation), it is reasonable to approximate
\begin{eqnarray}
P_{\nu+1} \approx P_1(e_m, U_m) \; P_{\nu}(U_{2N+1}, \ldots, U_{2N+\nu})
\end{eqnarray}
with $P_1(e_m,U_m)$ as the JPDF of  $U_m$ and $e_m$. 
Substitution of the above in eq.(\ref{qjk1})  leads to  
\begin{eqnarray}
Q_{jk} \approx 2N A_{jk} P_{\nu}
\label{qjk2}
\end{eqnarray}
 with $A_{jk}$ dependent only on non-zero eigenvalues and corresponding eigenfunctions:
\begin{eqnarray} 
 A_{jk}(Y) ={1\over 2N} \sum_{m=1}^{2N} \langle {U_{N+j,m} \; U^*_{N+k,m}  \over e_m^2}\rangle.
 \label{ajk}
 \end{eqnarray} 
As $A_{jk}=A_{kj}^*$,  the $(N+{\nu}) \times (N+{\nu})$ matrix $A = \left[ A_{kl} \right]$ is Hermitian.

Using eq.(\ref{qjk2}) in eq.(\ref{uzp}), the diffusion equation for $P_{\nu}$  now becomes
\begin{eqnarray}
{1\over v^2} \; {\partial P_{\nu} \over \partial Y} = 
 {\beta N } \; \sum_{n=2N+1}^{2N+\nu}\;  \sum_{k,j=N+1}^{2N+\nu} \left( A_{jk}\; {\partial^2 P_{\nu} \over \partial U_{jn} \partial U_{kn}^*} + A_{jk}^* \; {\partial^2 P_{\nu} \over \partial U_{jn}^* \partial U_{kn}} \right) 
\label{uzp1}
\end{eqnarray}
As clear from the  above, the components of the eigenfunctions $U_{2N+1}, \ldots, U_{2N+\nu}$ are mutually independent and their JPDF can be separated as follows:
\begin{eqnarray}
P_{\nu}(U_{2N+1},\ldots, U_{2N+\nu}) =\prod_{\eta=1}^{\nu} \;P_{1}(U_{2N+\eta}),
\label{vzp2}
\end{eqnarray}
where $P_{1}(U_{2N+\eta})$ is the JPDF of $N+\nu$ non-zero components of $U_{2N+\eta}$  (as $U_{k,2N+\eta}=0$ for all $k=1 \to N$), 
\begin{eqnarray}
P_{1}(U_{2N+\eta}) \equiv P_{1}(\{U_{N+k,2N+\eta}\}) \equiv P_{1}(U_{N+1,2N+\eta}, \ldots, U_{2N+\nu, 2N+\eta} ),
\label{vzp1}
\end{eqnarray}
For ease of presentation, hereafter we redefine symbols as $U_{N+k,2N+\eta} \to x_k$, $N+\nu \to N_{\nu}$ and $P_{1}(\{U_{N+k,2N+\eta}\}) \to P_{cs}(\{x_{k}\})$ with subscript $''cs''$ referring to  a chiral state. 

\subsubsection{Diffusion of a single chiral state}

Eq.(\ref{uzp1}) along with eq.(\ref{vzp2}) can now be used to derive the diffusion equation for $P_{cs}(U_{2N+\eta})$ for arbitrary $\eta$,
\begin{eqnarray}
{1\over v^2} \; {\partial P_{cs} \over \partial Y} = 
 {\beta N} \; \sum_{k,j=1}^{N_{\nu}} \left(A_{jk} \; {\partial^2 P_{cs} \over \partial x_j \partial x_k^*} +A_{jk}^* \; {\partial^2 P_{cs} \over \partial x_{j}^* \partial x_{k}} \right) 
\label{uzp2}
\end{eqnarray}
 As the ortho-normalization condition on $U_{2N+\eta}$ gives the condition $\sum_{k=1}^{N_{\nu}} |x_k|^2=1$ with ${1\over N_{\nu}} \le |x_{k}|^2 \le 1$, this subjects $P_{cs}$ to following boundary condition: $P_{cs}(\{x_k \}; Y) \to 0$ for $|x_{k}| \to 1$ for arbitrary $k$ and $x_l \not =0 \forall l \not=k$. Further $P_{cs}$  is also subjected to initial condition $P_{cs}(\{x_k \}; Y_0)$.

A general solution of the above equation can be obtained as follows. Consider $\mu =\left[ \mu_n \delta_{mn} \right]$ as the eigenvalue matrix and $V =\left[V_{kl} \right]$ as the eigenfunction matrix of $A$. As $A$ is Hermitian, we have $A=V. \mu V^{\dagger} $. Eq.(\ref{uzp2}) can now be simplified by following transformation 
\begin{eqnarray}
\sum_j V_{jn} {\partial \over \partial x_j} = {\partial \over \partial r_n}
\label{trx}
\end{eqnarray}
and
\begin{eqnarray}
P_{cs}(\{ x_k \}) ={\rm Det}(J) \; P_r(\{r_j\})
\label{prx}
\end{eqnarray}
 with $J(r|x)$ as the jacobian matrix of transformation from the set of variables $\{x_k \} \to \{r_n \}$: $J \equiv [\partial_k r_m]_{k,n =1\to N_{\nu}}$. Following from eq.(\ref{trx}), $\sum_k V_{kn} {\partial r_m\over \partial x_k} = \delta_{mn}$ or alternatively $V [\partial_k r_m] =1$ and thereby $J= V^{-1}$; this implies ${\rm Det}(J) ={1\over Det (V)}$. Further from eq.(\ref{trx}), we have 
\begin{eqnarray} 
 r_m=\sum_{m,n} V_{mn}^* \; x_n. 
\label{rx}
\end{eqnarray}
Using the above transformation, eq.(\ref{uzp2}) can be rewritten as  
\begin{eqnarray}
{1\over v^2 \beta N \; {\rm Det} (J) } \; {\partial \; ({\rm Det} (J) \; P_{r} )\over \partial Y} = 
 \sum_{n=1}^{N_{\nu}} \; \mu_n \;  {\partial^2 P_{r} \over \partial r_n \partial r_n^*}
\label{uze2}
\end{eqnarray}
The solution of the above equation can now be given as 
\begin{eqnarray}
P_{r}(\{r_k\}, Y) = \omega  \; {\rm e}^{- v^2 \beta N E (Y-Y_0) } \; \prod_{n, s=1}^{ N_{\nu}, \beta}   \cos\left[ k_{ns} \; r_{ns} + \theta_{ns}\right] 
\label{chw1}    
\end{eqnarray}
with notation $r_n=r_{n1}+i r_{n2}$, $\omega \equiv  {{\rm Det} V(Y) \over {\rm Det} V(Y_0)}$ and $E= \sum_{n=1}^{N_{\nu}} \sum_{s=1}^{\beta} \mu_n \; k_{ns}^2$ where $k_{ns}$ and $\theta_{ns}$ are arbitrary constants. This in turn leads to
\begin{eqnarray}
P_{cs}(\{x_k\}, Y) = C\;  {\rm e}^{- v^2 \beta N E (Y-Y_0)}
\;\prod_{n, s=1}^{ N_{\nu}, \beta} \cos \left(\Phi_{ns} +\theta_{ns} \right)
\label{chw2}    
\end{eqnarray}
with $C$ as a constant and 
\begin{eqnarray}
\Phi_{n1}  &\equiv&  k_{n1} \sum_{m=1}^{ N_{\nu}} \sum_{s=1}^{\beta} V_{nm;s} \; x_{ms}, \\
\Phi_{n2}  &\equiv&  k_{n2} \;\sum_{m=1}^{ N_{\nu}} \sum_{s\not=s'=1}^{\beta} (-1)^{s'} V_{nm;s} \; x_{ms'}.
\label{phic}
\end{eqnarray}

The constants $k_{ns}$ and $\theta_{ns}$ can be determined by the boundary conditions. As a simple example, here we consider the case with $H$ as a real-symmetric matrix ($\beta=1$). This in turn renders $U$ as an orthogonal and thereby $A$ and $V$ as real-symmetric and real matrices respectively. Choosing the solutions of definite parity along with boundary condition $P_{cs}(\{x_k\}, Y) \to 0$ for $x_{ks} \to \pm 1$ $\forall \; k$ then gives $\theta_n=0$ and $\Phi_n =  k_{n} \sum_{m=1}^{ N_{\nu}}  V_{nm} = {l_n \pi\over 2}$, with $l_n= \pm 1, \pm 2\ldots$ or 
$k_n = {\pi l_n  \over 2 \chi_n}$ with $\chi_n=\sum_{m=1}^{ N_{\nu}}  V_{nm}$; (as all variables are real in this case, we suppress the subscript $''1''$ from them).
As a consequence, $E$ can take many possible discrete values  $E_l=\sum_{n=1}^{N_{\nu}}  {\pi^2 \; l_n^2 \; \mu_n \over 2 \; \chi_n^2}$, corresponding to various combinations of $l \equiv \{l_1, l_2,\ldots, l_{N_{\nu}}\}$, which in turn  leads to many possible solutions for $P_{cs}(\{x_k\}, Y) $.
The general solution can now be written as 
\begin{eqnarray}
P_{cs}(\{x_k\}, Y) = \sum_{l} C_l  \; {\rm e}^{- v^2 \beta N E_l (Y-Y_0)}
\;\prod_{n=1}^{ N_{\nu}}   \cos \left( {\pi l_n  \over 2 \chi_n}\sum_{m=1}^{N_{\nu}}  V_{nm} x_m \right)
\label{chw3}    
\end{eqnarray}
The constants $C_l$ can now be determined from the initial condition $P_{cs}(\{x_k\}, Y_0)$ and using the relation $\sum_{n=1}^{\infty} \cos(nx) \cos(ny) = \delta(x-y)$: 
$$C_j = \int_0^1 P_{cs}(\{x_k \}, Y_0) \; \prod_{n=1}^{ N_{\nu}} \cos \left( {\pi j_n  \over 2 \chi_n}\sum_{m=1}^{N_{\nu}}  V_{nm} x_m \right) \; {\rm D}x$$.

\subsubsection{Local Intensity of a Chiral State}

The local intensity $P_t(t)$ of a chiral eigenfunction with components $x_k$ can be defined as

\begin{eqnarray}
P_{t}(t)={1\over N_{\nu}} \langle \sum_{k=1}^{N_{\nu}} \delta(t-|x_k|^2) \rangle 
=  {1\over N_{\nu}} \sum_{k=1}^{N_{\nu}} \int \delta(t-|x_k|^2) \; P_{cs}(\{x_k\};Y) \; {\rm D}x.
\label{pt}
\end{eqnarray}
with  ${1\over N_{\nu}} \le x_k \le 1 \forall k=1 \to N_{\nu}$ and ${\rm D}x \equiv \prod_{k=1}^{N_{\nu}} {\rm d}x_k$.

 Differentiating the latter with respect to $Y$, subsequent use of eq.(\ref{uzp2}), followed by repeated partial integration then leads to 

\begin{eqnarray}
 {\partial P_t \over \partial \Lambda_t} = {\partial  \over \partial t} \left( t {\partial P_t\over \partial t}   \right)
 \label{uzp4}
\end{eqnarray}
where $\Lambda_t =v^2 {\mu_0 \beta }  (Y-Y_0)$ with 
\begin{eqnarray}
\mu_0=\sum_{k=1}^{N_{\nu}}  A_{kk}(Y) = {1\over 2N} \sum_{m=1}^{2N} \langle {1  \over e_m^2}\rangle ={1\over 2} \int {R_1(e;Y)\over e^2} \; {\rm d}e.
\label{ajks1}
\end{eqnarray}
Here the $2nd$ equality follows from the relation $\sum_{k=1}^{N_{\nu}}|U_{N+k,m} |^2=1/2$.

As the   limit $Y \to \infty$, and thereby $\Lambda_t \to \infty$, corresponds to equilibrium state (i.e Ch-GOE, Ch-GUE for $\beta=1$ and $2$ respectively); the solution in this limit can be obtained by setting  ${\partial P_t \over \partial \Lambda_t}=0$ in eq.(\ref{uzp4}). This  leads to
\begin{eqnarray}
P(t;\infty) =C_1+ C_2 \log t 
\label{pt1}
\end{eqnarray}
 with $C_1, C_2$ as arbitrary integration constants.
The normalization condition $\int_{1/N_{\nu}}^1 P_t(t; \Lambda_t) \; {\rm d}t =1$ gives $C_1=C_2+1$. The boundary condition  $\lim_{t \to 1} P_t(t) \to 0$ gives $C_1=0$ and thereby $P(t) = - \log t$,  thus implying a diverging $P(t)$ as $t \to {1\over N}$ in large $N$ limit. 


With $Y$ absent from the right side of eq.(\ref{uzp4}), its general solution  for arbitrary initial condition, say at $Y=Y_0$, can be obtained by separation of variables approach. Based on boundary condition at $t=0$, two possible solutions,  for finite $\Lambda_t$ and of definite parity, can be given as follows.

{\it (i) Cases with finite $P_t(0,0)$:}  The solution turns out to be

\begin{eqnarray}
 P_t(t;\Lambda_t) =\sum_n C_n \; {\rm exp}(- E_n \; \Lambda_t)   \; J_0(2\sqrt{E_n t}) 
\label{ptc1}
\end{eqnarray}
where $E_n= z_n^2/4$ with $z_n$ as the $nth$ zero of $J_0$, the Bessel function of first kind, and 
\begin{eqnarray}
 C_n= {2 \over J_1^2(z_n)} \; \int_0^1 P_0(x) \; J_0(z_n \sqrt{x}) \; {\rm d}x 
\label{ptc2}
\end{eqnarray}

{\it (ii) Cases with diverging $P_t(0,0)$:}  The solution  in this case is

\begin{eqnarray}
 P_t(t;\Lambda_t) =\sum_n C_n \; {\rm exp}(- E_n \; \Lambda_t)   \; N_0(2\sqrt{E_n t}). 
\label{ptc3}
\end{eqnarray}
Here again $E_n= z_n^2/4$ but now $z_n$  refers to $nth$ zero of $N_0$, Bessel function  of the 2nd kind. However as $N_0$ does not have orthogonality relations, the determination of coefficients $C_n$ for this case is no longer straighforward (see {\it appendix F}).

\subsubsection{Generalized Inverse Participation Ratio}

The localization behavior of chiral states has evoked a lot of interest in context of disordered bipartite lattices; the existing, system specific results indicate the state to be neither localized or delocalized, size-independent for all disorders  with critical characteristics. This motivates us to pursue a similar query in context of the Brownian ensemble. The standard route to analyze this behavior is through the moments of the eigenfunction intensity $I_{n}$, also referred as the  generalized inverse participation ratio: $I_{n}  =\sum_{k=1}^{N_{\nu}} |x_k |^{2n}  $. Its ensemble average can be given in trems of $P_t(t)$:  $\langle {\mathcal I}_{n} \rangle = \int t^n \; P_t(t) \; {\rm d}t = \langle t^n \rangle$.  Multiplication of eq.(\ref{uzp4}) with $t^n$ followed by the integration then gives a hierarchical equation for the moments of $P_t(t;\Lambda_t)$: ${\partial \langle I_{n} \rangle \over \partial \Lambda_t} = n^2 \; \langle I_{n-1} \rangle$.

Alternatively  the moments for the limit $\Lambda_t \to \infty$ can be derived directly from eq.(\ref{pt1}),
\begin{eqnarray}
 \langle I_n \rangle =\langle t^n \rangle = {1\over (n+1)^2} \left(C_1 (n+1)-C_2 \right) +O \left({1\over N^{n+1}} \right) \approx  {1\over (n+1)^2} 
\label{tp2}
\end{eqnarray}
Noting that $\langle I_n \rangle \sim N^{-n}$ in case of an eigenstate delocalized in the whole basis space and $\langle I_n \rangle \sim \xi^{-n}$ 
for the localized eigenstate with average localization length $\xi$ \cite{mj}, eq.(\ref{tp2}) suggests the chiral state to be different from both cases. 

Similarly, the moments for finite $\Lambda_t$ can also be obtained. For example, for case (i) (eq.(\ref{ptc1})) mentioned above, $ \langle I_n \rangle$ can be given as 
 \begin{eqnarray}
 \langle I_n \rangle &=&\langle t^n \rangle =\sum_n C_n \; {\rm exp}(- z_n^2 \; \Lambda_t /4 )   \; {\mathcal I}_n 
 \label{tp3}
 \end{eqnarray}
with
\begin{eqnarray}
 {\mathcal I}_n &=& n! \; \left( _pF_q \left[{n+1},{n+2,1},-{z_n^2\over 4}\right] - N^{1-n} \;  _pF_q \left[{n+1},{n+2,1},-{z_n^2\over 4N}\right]\right)
 \label{tp4}
 \end{eqnarray}
and $C_n$ given by eq.(\ref{ptc2}).

\section{Spectral Fluctuations}

As in the case of eigenfunctions, an integration of  eq.(\ref{ch19}) over undesired eigenvalues leads to  diffusion equations for the probability densities of remaining eigenvalues of $H$.  As discussed below, the diffusion of a single eigenvalue density occurs at a different scale from those of higher order ones and it is appropriate to consider them separately. 

\subsection{Level density}

The $Y$-dependent ensemble averaged level density $R_1(e; Y)$ can be defined as 
\begin{eqnarray}
R_{1,ch}(e; Y)=\sum_{n=1}^{2N+\nu} \langle \delta(e-e_n) \rangle =\sum_{n=1}^{2N+\nu} \int \delta(e-e_n) \; P_{ev}(e_1, e_2,\ldots, e_{2N+\nu}; Y) \; {\rm D}E.
\label{rr0}
\end{eqnarray}  
where ${\rm D}E=\prod_{n=1}^{2N+\nu} {\rm d}e_n$ with $-\infty \le e_n \le \infty$.
Using the definition (\ref{rr0}), the level density $R_{1,ch}(e;Y)$ in chiral case can be expressed in terms  of the level density $R_{1}(e;Y)$ of the eigenvalues $e_1, \ldots, e_N$
\begin{eqnarray}
R_{1,ch}(e;Y)=  R_{1}(e;Y) + R_{1}(-e;Y) + \nu \; \langle \delta(e) \rangle
\label{rd0}
\end{eqnarray} 
where
\begin{eqnarray}
R_{1}(e; Y)=\sum_{n=1}^{N} \langle \delta(e-e_n) \rangle =\sum_{n=1}^{N} \int_0^{\infty} \delta(e-e_n) \; P_{N}(e_1,\ldots, e_N; Y) \; \prod_{n=1}^{N} {\rm d}e_n.
\label{rr0a}
\end{eqnarray}  
with $P_N$ defined in eq.(\ref{pe1}). Here $R_1(e;Y)$ is subjected to the normalization condition : $\int_0^{\infty} R_1(e) \; {\rm d}e = N$

The above definition along with a direct integration of eq.(\ref{ch19}) over $N-1$ eigenvalues and entire eigenvector space leads to an evolution equation for $R_1(e;Y)$, from an arbitrary initial condition $R_1(e;Y_0)$,

\begin{eqnarray}
{\partial R_{1} \over\partial Y} &=& {2 \beta}\; {\partial \over \partial e} \left( \gamma \; e  -{{\nu+1/2}\over e}- {\bf P}\int_{0}^{\infty}   {2 \; e \; R_{1}(e') \over e^2-e'^2} \; {\rm d}e' \right) \; R_{1} + {\partial^2  R_{1} \over \partial e^2}
\label{r1}
\end{eqnarray}
As clear from the above,  $R_{1}(e)=R_{1}(-e)$.  The evolution occurs at a  scale $Y \sim N \Delta_e^2$ with $\Delta_e(e)$ as the local mean level spacing in a small energy-range around $e$. 

\subsubsection{Behavior for non-zero energies}
For regions $|e| \gg 0$, the  drift term with $e^{-1}$ as well as the diffusion term can be neglected (being of $O(1/N)$ with respect to other terms). Writing $\int_{0}^{\infty}   {2 \; e \; R_1(e') \over e^2-e'^2} \; {\rm d}e'   = \int_{-\infty}^{\infty}   {R_1(e') \over e-e'} \; {\rm d}e'$, eq.(\ref{r1}) can be approximated as 
\begin{eqnarray}
{\partial R_{1} \over\partial Y} &=& {2 \beta} \; {\partial \over \partial e} \left( \gamma \; e  -  {\bf P}\int_{-\infty}^{\infty}   {R_{1}(e') \over e-e'} \; {\rm d}e'  \right) \; R_{1}(e) 
\label{rr1}
\end{eqnarray}

Referred as Dyson-Pastur equation, the above equation is analogous to that 
 for the level density of a non-chiral Brownian ensemble; the solution for former can then be obtained from already known solutions for the latter.  As discussed in \cite{sp} (see eq.(33) and appendix D therein),  the solution of eq.(\ref{rr1}) can be given as 
 \begin{eqnarray}
 R_1(e; Y)={1\over \pi} \; \lim_{\varepsilon \to 0} \; G(e-i  \varepsilon; Y)
 \label{r1ne}
 \end{eqnarray}
where $G(z;Y)$ is the resolvent, $G(z; Y) =\int   {R_1(x) \over z-x} \; {\rm d}x$, and satisfies following differential equation:
${\partial G \over\partial Y} = {2 \beta} \; {\partial \over \partial e} \left(z G(z;Y)- {1\over 2} G^2(z;Y)\right)$. The solution of the latter can be given as $G(z;Y)=G(z-Y G(z; Y); Y_0)$ \cite{app}; (using this approach, the derivation of $R_1(e;Y)$   for the case with an initial Gaussian level density is also discussed in \cite{psf}).

The limit $Y \to \infty$ corresponds to a semi-circle level density (expected as $\rho(H)$ in eq.(\ref{rhoh}) approaches Ch-GOE/ Ch-GUE in the limit) but this limit is never reached if the initial condition $R_1(e; Y_0)= \delta(e)$ \cite{apps}.

\subsubsection{Behavior near zero energy}
Near $e \sim 0$, eq.(\ref{r1}) deviates significantly from its non-chiral counterpart. This is because the singular term $ 1/e$  in eq.(\ref{r1}) now dominates which makes it necessary to retain  the diffusion term too.
Our interest is in the solution of eq.(\ref{r1}) near $ e\sim 0$. With left and right side of the above equation dependent on different variables i.e $Y$ and $e$, respectively, it is instructive to consider separation of variables approach and consider 

\begin{eqnarray}
R_1(e,Y) ={\mathcal N_q} \; {\rm exp}(-q(Y-Y_0)) \; f_q(e) 
\label{nr1}
\end{eqnarray}
with $f_q(e)$ as an arbitrary  function of $e$ and ${\mathcal N_q}, q$ as an arbitrary positive constant. 
Substitution of  the above in eq.(\ref{r1}), leads to 
\begin{eqnarray}
  {\partial^2 f_q\over \partial e^2} +  {b_1(e)\over e} \;  {\partial f_q \over \partial e} + {b_0(e)\over e^2} \; f_q= 0
\label{rr3}
\end{eqnarray}
where 
\begin{eqnarray}
b_0(e) &=& (1/4)(\beta (2\nu+1) + 2(\beta  {\partial \alpha \over \partial e} + q) e^2), 
\label{b0}\\
b_1(e) &=& (1/4)\beta (2\alpha \; e-(2\nu+1)), \\
\label{b1}
\alpha(e;Y) &=&  \gamma \; e -{\bf P} \int_{-\infty}^{\infty}   { R_1(e',Y)  \over e-e'} \; {\rm d}e'
\label{alpy}
\end{eqnarray}
 With $R_1$ as a probability density, $f_q$ is required to be positive semidefinite. Further the unknown constants ${\mathcal N_q}$ and $q$ can be determined by the boundary conditions and known initial condition $R_1(e; Y_0) = {\mathcal N_q} \; f_q(e) $.
 
 The next step depends on the behaviour of $\alpha(e)$ near $e \sim 0$.
 As clear from eq.(\ref{alpy}), $\alpha(e)$ depends on the solution itself. Although, for a chirality preserving transition, it is intuitively encouraging to assume $R_1(e;Y) \sim R_1(e; Y_0)$ near $e\sim 0$  but the spectral correlations for $|e| >0$ are expected to evolve with varying $Y$ and it is not a priori obvious whether it could affect the behavior near $e=0$. 

An insight in $R_1(e \sim 0;Y)$-behavior can however be gained by approximating the integral in  eq.(\ref{alpy}) by a power law near $e \sim 0$.
 Assuming $\lim_{e \to 0} \; \alpha(e) \sim -\gamma \; e + c_0 \; {\rm e}^{\mu}$ with $c_0$ as a constant and $\mu$ arbitrary, 
$b_0, b_1$ can be approximated by their  $e \to 0$ limits. Writing the general solution as 
\begin{eqnarray}
f_q(e)=C_1 \; f_1(e) + C_2 \; f_2(e),
\label{fte}
\end{eqnarray}
the forms of $f_1(e), f_2(e)$ depends on $\mu$  and are given in Table 1 for some $\mu$-values. Here $C_1, C_2$ are arbitrary constants to be determined by the boundary conditions on $f_q(e)$. From eq.(\ref{nr1}), the existence of $R_1(e, Y)$ in the limit $Y \to \infty$ requires $q=0$. But as $b_0(e), b_1(e)$ near $e \sim 0$ are $q$-independent, none of the solutions of eq.(\ref{rr3}) depend on $q$.  This implies no change in the $R_1(e; Y)$ behavior near $e \sim 0$ as $Y$ varies
\begin{eqnarray}
\lim_{e \to 0} R_1(e;Y) \approx \lim_{e \to 0} R_1(e;Y_0) =C_1 \; f_1(e) + C_2 \; f_2(e).
\label{fte1}
\end{eqnarray}
As an example, consider the case $R_1(e,Y_0) = {1\over \sqrt{2 \pi}} \; {\rm exp}[-e^2]$ which implies $R_1(e \sim 0,Y_0) \to {\rm constant}$. 
With ${\bf P} \int_{-\infty}^{\infty}   { R_1(e',Y_0)  \over e-e'} \; {\rm d}e' \approx 0$,  this gives $\alpha(e; Y_0)=-\gamma e$. Approximating $\alpha(e;Y_0)$ by $\alpha(e; Y)$ in eq.(\ref{rr3}), the solution now corresponds to case 2 of the table 1 with $c_0=0$. For $\gamma=1/2$,  this gives $R_1(e \sim 0,Y) \to {\rm constant}$ which is consistent with numerical result given in \cite{psmulti}.

Eq.(\ref{fte1}) describes the general solution of eq.(\ref{r1}) near $e \sim 0$. 
It is worth noting, for later reference,  a particular solution  for eq.(\ref{r1}) near $e \sim 0$ can be given as
\begin{eqnarray}
R_1(e, Y)  \approx {\mathcal N_e}  \; e^{-|\eta|} \; {\rm exp}\left[-\Omega (-\log e)^b \right]
\label{rr4}
\end{eqnarray}
As the above solution  is obtained by neglecting terms $(-\log e)^{b-2}$ and $(-\log e)^{2(b-1)}$ in comparison with $(-\log e)^{b-1}$, it is valid only for $0 < b <1$. Here the constant $\eta$ and and $\mathcal N_e$ depends on the initial condition at $Y_0$ and the normalization condition on $R_1$ respectively.

\begin{table}[h!]
  \begin{center}
\centering
\caption{\textbf{Behaviour of $R_1(e;Y)$ near $e \sim 0$:} Here the column $2nd$ contains  various power law dependence of $\alpha(e \sim 0)$ and the columns $3rd$ and $4th$ corresponding $b_0(e)$ and $b_1(e)$, with $\alpha(e)=-\gamma e +c_0 \; e^{\mu}$, $a=-(2\nu+1) \beta/4$, $g=c_0 \beta/2$, $s=?$. The two corresponding solutions of eq.(\ref{rr3r})  are given in $6th$ column; note for cases $3-6$, the solutions mentioned are  only for specific $\mu$-value given in column $5th$).  As clear from  columns $3rd$ and $4th$, $b_0, b_1$ for none of the cases depend on $q$. This in turn implies $R_1(e, Y) \approx R_1(e,Y_0)$ and is therfore consistent with our conjecture.} 
\begin{tabular}{|c|c|c|c|c|p{10cm}|}
\hline
Index & $\alpha(e)$ & $b_1(e)$ & $b_0(e)$ & & $f_1(e), f_2(e)$  \\
\hline
1. &$-\gamma e+c_0 e^{\mu}$ &a & $-(a+\beta \gamma/2)$ & $\mu >1$  & $f_1=e^{{1\over 2}[(1-a)-\sqrt{(1-a)^2+4 (a+\beta \gamma/2)}]}$\\
& & & & & $f_2=e^{{1\over 2}[(1-a) +\sqrt{(1-a)^2 + (a+\beta \gamma/2)}]}$ \\ \hline

2. & $-\gamma e+c_0$	&  a & $-(a+\beta \gamma/2)$ & $\mu=0$ &  same as case 1 \\\hline

3. & $-\gamma e+c_0 e^{\mu}$ &a & $\mu \; g \; e^{\mu-3}$& $\mu={1\over 2}$  & $f_1=\left({2 g \over \sqrt{e}} \right)^{a-1} \; J_{2(1-a)} \left(2\left({4 g^2\over e}\right)^{1\over 4} \right)  \Gamma(3-2a)$\\ 
& & & & & $f_2=\left({2 g \over \sqrt{e}} \right)^{a-1} J_{-2(1-a)} 
\left(2 \left({4 g^2\over e}\right)^{1\over 4} \right) \Gamma(2a-1)$, 
\\\hline

5. & $-\gamma e+{c_0 \over e^{\mu}}$ & $a$ & $- {\mu \; g \over e^{\mu+3}} $ & $\mu={1\over 2}$    & $f_1=\left({2 g \over 9 \sqrt{e^3}} \right)^{a-1\over 3} \; I_{-2(a-1)\over 3}
\left({2\over 3} \left({4 g^2\over e^3}\right)^{1\over 4} \right) 
 \Gamma\left({5-2a \over 3} \right)$ \\
 & & & &  &$f_2=(-1)^{2(a-1)\over 3} \left({2 g \over 9 \sqrt{e^3}} \right)^{a-1\over 3} \; 
I_{2(a-1)\over 3}\left({2\over 3}\left({4g^2\over e^3}\right)^{1\over 4} \right)  \Gamma\left({1+2a \over 3} \right)$\\ \hline

6. & $-\gamma e +{c_0 \over e}$ & $g+a$ & $- {g\over e^{4}}$ & $\mu=1$ &
$f_1={1\over \sqrt{\pi}} \left({2 \sqrt{g} \over e} \right)^{-s\over 2} \; {1\over x} \; K_{s\over 2}
\left({2\over 3}\right) \left({ \sqrt{g}\over e}\right)$ \\
 & & &&  &$f_2={1\over x} \; {\rm exp}\left[- {\sqrt{g} \over e}\right]   \; 
F\left({1\over 2}(1+s), 1+s, {2 \sqrt{g}\over e}\right)$, $s=a+g-1$\\ \hline

7. & $-\gamma e +\log e $ &a & ${g \over e^3}$ &  & $f_1= \left({g \over e} \right)^{(a-1)/2} \; 
J_{-(a-1)}\left(2 \sqrt{g\over e} \right) \Gamma(2-a)$ \\
& & & & & $f_2=\left({g \over e} \right)^{(a-1)/2} \; J_{a-1}\left(2 \sqrt{g\over e} \right)  \Gamma(a)$\\ \hline
\hline
\end{tabular}      
\label{ae}
\end{center}
\end{table}

\subsection{Static Correlations}

 The $n$-level correlation $R_{n,ch}(e_1,...,e_n;Y)$ i.e. the 
probability density for $n$ levels  to be at $e_k > 0$, $k=1 \to n$, irrespective  of the position of other $N-n$ levels, can be defined as 
\begin{eqnarray}
R_{n,ch}(e_1,e_2,..,e_n;Y)={ N! \over {(N-n)!}} \; \int P_{ev}(E;Y) \; \prod_{k=n+1}^{2N+\nu} {\rm d}e_{k}. 
\label{rd1}
\end{eqnarray}
Note here the term  ${ N! \over {(N-n)!}} $ corresponds to consideration of only $N$  positive definite levels; the latter, once chosen, pin the $N$ negative definite levels. The above gives 
\begin{eqnarray}
R_{N,ch}(e_1,e_2,..,e_N;Y) &=& N! P_N(e_1,e_2,..,e_N;Y)  \\
R_{n,ch}(e_1,..,e_n;Y) &=& R_{n}(e_1,..,e_n;Y) 
\label{rd2}
\end{eqnarray}
where $R_{n}(e_1,..,e_n;Y) ={ N! \over {(N-n)!}} \; \int P_{N}(e_1,\ldots,e_N;Y) \; \prod_{k=n+1}^{N} {\rm d}e_{k}$.
  
The $Y$-governed diffusion equation for $R_n$ can be obtained by first differentiating eq.(\ref{rd1}) with respect to $Y$ and subsequently using eq.(\ref{ch19}). 
Similar to Hermitian case without chirality, here again the evolution of $R_n$  occurs on the scales determined by $Y-Y_0 \sim \Delta_e(e)^2$ with $\Delta_e(e)$ as the local mean level spacing at energy $e$ \cite{apbe,sp,fkpt}; the transition in $R_n$ and therefore other spectral fluctuation measures are governed by the rescaled parameter    
\begin{eqnarray}
\Lambda_e(Y,e)={ (Y-Y_0) \over \Delta_e^2}.
\label{alm1}
\end{eqnarray}
The $\Lambda_e$-governed diffusion of rescaled correlations ${\mathcal R}_n(r_1,..,r_n;\Lambda_e)= \lim_{ N\rightarrow \infty} \;\Delta_e^n \; {\it R}_n(e_1,..,e_n; Y)$, with $r_n = {(e_n-e)\over \Delta_e}$ and $e=r \; \Delta_e$, from an  arbitrary initial condition, can be given as 

\begin{eqnarray}
{\partial {\mathcal R}_n \over\partial \Lambda_e} = \sum_{j=1}^n  {\partial \over \partial r_j} \left[ {\partial {\mathcal R}_n\over \partial r_j}- \sum_{k=1;\not=j}^n {  2\; \beta \; (r_j+r) \; {\mathcal R}_n \over {(r_j-r_k)(r_j + r_k+2 r)}} -   {\beta (\nu+1/2) \; {\mathcal R}_n\over (r_j+r)} + 2 \beta \int_{}^{} {  (r_j+r) \; {\mathcal R}_{n+1} \over {(r_j+r)^2-t^2}} \; {\rm d}t\right]. \nonumber \\
\label{rn}
\end{eqnarray}
with ${\mathcal R}_{n+1} \equiv {\mathcal R}_{n+1}(r_1,\ldots,r_n, t; \Lambda_e)$. 
Here the limits $\Lambda_e=0, \infty$ correspond to the initial and the stationary state for the $R_n$, respectively.   The solution of eq.(\ref{rn}) in the stationary limit  corresponds to Ch-GOE/ Ch-GUE and can be obtained by substituting $  {\partial R_n \over \partial \Lambda_e}=0$ in eq.(\ref{rn}).

As a  next step, it is desirable to solve the above equation. But  the general solution of a similar equation is not available for the non-chiral Brownian ensembles too.  A number of insights about the statistics can however be gained without a detailed solution. For example, 

(i) eq.(\ref{rn}) depends on the scale $e$ of the local correlations; this is  contrary to  non-chiral Hermitian case (see eq.(16) of \cite{apbe}), 

(ii) the solution of eq.(\ref{rn}) for any non-zero, finite $\Lambda_e$ corresponds to an intermediate, non-equilibrium correlation among $n$-levels in a neighbourhood of energy $e$ within a range in which $R_1(e)$ can be assumed to be constant. But as $R_1$ itself is $Y$-dependent, the local statistics can vary significantly along the spectrum,   

(iii) For ${\mathcal R}_n(r_1,\ldots, r_n)$  with all $r_1,\ldots, r_n >0$ (or for all $r_1,\ldots, r_n <0$),  eq.(\ref{rn}) can be approximated by one similar to that of Hermitian case without chirality (eq.(16) of \cite{apbe}, also see \cite{sp} for a detailed discussion)

\begin{eqnarray}
{\partial {\mathcal R}_n \over\partial \Lambda_e} &=& \sum_j {\partial^2 {\mathcal R}_n\over \partial r_j^2}-2\; \beta \sum_{j\not=k} {\partial \over \partial r_j} \left({  {\mathcal R}_n \over {r_j-r_k}}\right) -2 \; \beta \sum_j {\partial \over \partial r_j} \int_{}^{} {{\mathcal R}_{n+1} \over {r_j-t}} \; {\rm d}t. 
\label{rn1}
\end{eqnarray}
As a consequence, for regions away from $e \sim 0$, the local statistics for the Hermitian case with chirality is almost analogous to the one without chirality; ${\mathcal R}_2$ for the latter case for many initial conditions is  given in \cite{apbe} and  can directly be used for the chiral cases with same $\Lambda_e$ and  analogous initial conditions. (Note, the fluctuations for the case with $\nu=0$ and for a GUE initial state are discussed in \cite{km1}).   

(iv) For $r=0$ i.e  $e=0$, the $3rd$ term inside square bracket in the right side of eq.(\ref{rn}) dominates over the $2nd$ term. Keeping only the dominant terms, eq.(\ref{rn}) can be approximated as 

\begin{eqnarray}
{\partial {\mathcal R}_n \over\partial \Lambda_e} = \sum_{j=1}^n  {\partial \over \partial r_j} \left[ {\partial {\mathcal R}_n\over \partial r_j}-   {\beta (\nu+1/2) \; {\mathcal R}_n\over r_j} + 2  \beta \int_{}^{} {  r_j \; {\mathcal R}_{n+1} \over {r_j^2-t^2}} \; {\rm d}t\right].
\label{rnz}
\end{eqnarray}
With interaction term now neglected, substitution of ${\mathcal R}_n(r_1,\ldots, r_n) = \prod_{k=1}^N {\mathcal R}_1(r_k)$ in the above equation  leads to eq.(\ref{r1}) with $e \to r_k$ and  $R_1(e) \to {\mathcal R}_1(r_k)$ which can again be solved as discussed in section VII. A. This suggests a weak corrrelations among energy levels near $e=0$ and thereby a spectral statistics different from Ch-GOE/ Ch-GUE.

\section{Connections}
The underlying complexity e.g. many body interactions in physical systems often manifests in the form of mathematical complexity. A knowledge of connections among different complex systems is therefore not only of fundamental relevance but can also help in technical insights about a system based on the available information for the other. Here we discuss the connections of chiral Brownian ensemble with a few other relevant complex systems.
  
\subsection{Calogero-Sutherland Hamiltonian (CSH)}

Eq.(\ref{ch19})  can 
be rewritten in terms of the Schrodinger equation for the interacting particles moving along a real line, subjected to confining potential;  the eigenvalues playing the role of 
the particles evolve with respect to an imaginary  time. With $|Q_N| = \prod_{n=1}^N e_n^{\beta(\nu+1/2)}  \; \prod_{j > k}^N |e_j^2 - e_k^2|^{\beta} \; {\rm e}^{-{\gamma \over 2\; v^2} \sum_{j=1}^N e^2_j }$, this follows by applying the transformation $\Psi=P_{N} /|Q_N|^{\beta/2}$   to eq.(\ref{ch19})  which reduces it as 
 \begin{eqnarray}
 {\partial \Psi \over\partial Y} = {\mathcal H} \; \Psi
 \label{cs1}
 \end{eqnarray}
where the 'Hamiltonian' ${\mathcal H}$ turns out to be a variant of the
CS Hamiltonian for $N$ fermions confined along a real line:
\begin{eqnarray}
{\mathcal H} =\sum_i  {\partial^2 \over \partial e_i^2} -
 {\beta(\beta-2)\over 4} \sum_{i,j; i\not=j}  \left({1\over (e_i-e_j)^2} + {1\over (e_i+e_j)^2}\right) + g_0 \sum_{k} {1\over e_k^2} 
 -{\gamma^2 \over 4}\sum_i  e_i^2 + c_0 \nonumber \\
 \label{h2}
\end{eqnarray}
with $c_0={N\over 2} (\alpha+1)\gamma + {\gamma \over 2}\beta N(N-1)$,  $g_0={\alpha \over 4}(2-\alpha)$ and $\alpha={\beta\over 2}(1+2\nu)$.

The "state" $\psi$ or $P_N(E,Y|E_0,Y_0)$ can formally
be expressed as a sum over the  eigenvalues $\lambda_k$ and eigenfunctions  $\psi_k$ of ${\mathcal H}$
\begin{eqnarray}
P_N(E,Y|E_0,Y_0) = {|Q_N(X)|\over |Q_N(Y)|}^{\beta/2} \; \sum_k {\rm exp}(-Y \lambda_k) \psi_k(Y) \psi_k(Y_0)
\label{p2}
\end{eqnarray}
As clear from the above, for $Y\rightarrow \infty$, the particles are in their ground state
$\psi_0 =  C_{\beta}^{1/2} \; |Q_N|^{\beta/2}$ 
 with a distribution $\psi_0^2$; note $\psi_0$ gives the correct 
form i.e eq.(\ref{stdis1}) for $P_N(Y\rightarrow \infty)$. An integration of eq.(\ref{p2}) over  the
initial state $P_N(E_0,Y_0)$  lead to the joint probability distribution $P_N(E,Y)=\int P_N(E,Y|E_0,Y_0) \; P_N(E_0, Y_0) \; {\rm d}e_{01}\ldots {\rm d}e_{0N}$ and
thereby static (at a single parameter value) density correlations $R_n$.

To proceed further, it is imperative to determine the eigenstates and the eigenvalues
of the Hamiltonian ${\mathcal H}$ which requires a detailed study. Some insight however can be gained by noting that for the case with $e_1, e_2, \ldots e_N >0$, ${\mathcal H}$  in eq.(\ref{h2}) can locally be approximated as the standard CS Hamiltonian ${\mathcal H}_{CS} $:
\begin{eqnarray}
 {\mathcal H}_{CS} = \sum_i  {\partial^2 \over \partial e_i^2} -
 {\beta(\beta-2)\over 4} \sum_{i,j; i\not=j}  {1\over (e_i-e_j)^2} 
 -{\gamma^2 \over 4}\sum_i  e_i^2 + c_0.
 \label{h3}
\end{eqnarray}
The local particle correlations for $ {\mathcal H}$ can then be approximated by ${\mathcal H}_{CS} $. The above also implies that the local correlations,  away from zero energy, of  chiral Brownian ensembles are analogous to those of  non-chiral Brownian ensembles (with Hermiticity as the only matrix constrains) with same $\beta$ (see also \cite{apbe, psbe} for more information about the latter). This is also consistent with the insights based on hierarchical equations for the  level correlations discussed in section VII B. 
Also note that, similar to case for the non-chiral Brownian ensembles \cite{sp, pcalo, ps-all, psnh}, here again the inverse square interaction term drops out for $\beta=2$, leaving both ${\mathcal H}$ and ${\mathcal H}_{CS}$ as free particle Hamiltonians but subjected to different confining potentials.  

The Hamiltonian in eq.(\ref{h3}) has been studied in great detail in past and many of its states and particle correlation have been worked out \cite{sp, apps}. The information can then be used in deriving the spectral correlations for the present case. (Although the steps are essentially same as applied in the case of Brownian ensembles with Hermitian condition \cite{sp, apps} but the difference in confining potential in eq.(\ref{h2}) and eq.(\ref{h3}) may manifest in the long range correlations). 
. 

 The explicit analysis of correlations near zero energy, involves technical handling of  various integrals and a separate study.  The present analysis reveals however an important
connection: the level correlations of different complex systems need not be studied
separately, a thorough probing of the particle correlations of
CS type Hamiltonian often  gives all the required information \cite{pcalo}; 
(see also \cite{psnh} in this context). The CS system being
integrable in nature, the semiclassical techniques can also be very successful for the probing.

\subsection{Wishart Brownian Ensembles}

Consider  $N \times N$  matrix $L_1 \equiv C.C^{\dagger}$ and $(N+\nu) \times (N+\nu)$ matrix $L_2 \equiv C^{\dagger}. C$ with $C$ same as in eq.(\ref{ch1}). Both $L_1$ and $L_2$ are known as Wishart matrices. Following from eq.(\ref{wis1}), we have
$L_1 \; X_n =   \lambda_n \; X_n$ and $ L_2 \; Z_n = \lambda_n \; Z_n$ where $\lambda_n=e_n^2$, $n=1 \to N$ are non-zero eigenvalues of $L_1$ and $L_2$ corresponding to eigenfunctions $X_n$ and $Z_n$, respectively; note $L_2$ also has $\nu$ zero eigenvalues.

The ensemble densities, say $\rho_{L_1}(L_1)$ and $\rho_{L_2}(L_2)$ of $L \equiv L_1, L_2$ can be expressed as $\rho_{L_1}(L_1) = \int \delta(L_1 - C.C^{\dagger}) \; \rho_c(C) \; {\rm D}C$ and $\rho_{L_2}(L_2) = \int \delta(L_1 - C^{\dagger}.C) \; \rho_c(C) \; {\rm D}C$. Based on  $\rho_c(C)$, $\rho_{L}$ can be of various forms e.g representing a Wishart stationary ensemble if $\rho_c(C )$ is stationary (free of parameters) and  non-stationary e.g. Wishart Brownian or multi-parametric ensemble if  $\rho_c(C )$ is non-stationary (e.g. see \cite{psmulti}).

As clear from the above, a chiral matrix $H$ and a Wishart matrix $L$  can both be written in terms of a complex matrix $C$. Eq.(\ref{ch8}) and eq.(\ref{chi8}) in the present paper and eq.(5) and eq.(7) in Ref.[38] describe the dynamics of a same matrix  C, and therefore, are same.  But the  appearance of $C$ in the matrix structures of $H$ and $L$ is different which in turn leads to two different type of ensembles. As expected, C being the basic component of both, the properties of the ensemble of chiral matrices are connected to Wishart ensemble; the  connection can be  elucidated as follows.

With $\lambda_n=e_n^2$, the JPDF $P_{\lambda}(\lambda_1,\ldots, \lambda_N)$ of the non-zero eigenvalues of $L_1$ and $L_2$  is related to the JPDF  $P_s(e_1, e_2, \ldots,  e_N)$ of the singular values of $C$ and $C^{\dagger}$.
For example, the jpdf  for $L_1, L_2$ described by a stationary Wishart ensemble is \cite{sp}
\begin{eqnarray}
P_{\lambda}(\lambda_1, \lambda_2, \ldots,  \lambda_N) &=& {\mathcal N}_w \; \prod_{n=1}^N \lambda_n^{(\nu+1)\beta/2-1 }  \; \prod_{j > k}^N |\lambda_j - \lambda_k|^{\beta} \; {\rm exp} \left[{-{(\gamma/ 2 v^2)} \sum_{j=1}^N \lambda_j }\right].
\label{stdis2}
\end{eqnarray}
with ${\mathcal N}_w$ as a normalization constant. The relation $\lambda_n=e_n^2$ then 
gives the JPDF  $P_s(e_1, e_2, \ldots,  e_N)$ of the singular eigenvalues of $C$ as 
\begin{eqnarray}
P_s(e_1, e_2, \ldots,  e_N) &=&  2^N 
\; \left(\prod_{n=1}^N \sqrt{\lambda_n}\right) \;\;  P_{\lambda}(\lambda_1, \ldots, \lambda_N)\label{pl1}\\
&=& 2^N \; {\mathcal N}_w \; \prod_{n=1}^N e_n^{(\nu+1)\beta-1 }  \; \prod_{j > k}^N |e_j^2 - e_k^2|^{\beta} \; {\rm exp}\left[{-{(\gamma/ 2 v^2)} \sum_{j=1}^N e^2_j }\right].
\label{stdis3}
\end{eqnarray}

It is worth noting here that $P_s(e_1, e_2, \ldots,  e_N)$ differs from $P_N(e_1, e_2, \ldots,  e_N)$ that appears in eq.(\ref{stdis1});  the additional factor of $e_n^{1-\beta/2}$ in the latter
comes from the correlations between equal and opposite pairs of the eigenvalues. This can also be seen from the non-stationary cases: for example, in case of a Wishart Brownian ensemble (WBE), the diffusion  of $P_{\lambda}$ can be described as \cite{sp, pslg}
\begin{eqnarray}
{1\over 4 v^2} \; \frac{\partial P_{\lambda}}{\partial Y}=  \sum_{n=1}^N \frac{\partial}{\partial \lambda_n}\left[\frac{\partial (\lambda_n \; P_{\lambda})}{\partial \lambda_n} - \left( \sum_{m=1}^N \frac{\beta \; \lambda_n}{\lambda_n- \lambda_m} + {\beta \; (\nu+1)\over 2} - {\gamma\over 2 v^2} \; \lambda_n \right)  P_{\lambda}\right]
\nonumber\\
\label{pdl1}
\end{eqnarray}
(Note the above equation corresponds to eq.(13) with rhs given by eq.(22 ) in \cite{sp} with $\tau=2 Y$, $v^2=1/4$ and $\gamma=1$; also note that eq.(22)  in \cite{sp}  has a prefactor $''1/2''$ missing). 
Here again a substitution of $\lambda_n =e_n^2$ in the above leads to a diffusion equation for $P_s$; the equation turns out to be same as eq.(\ref{ch19}) but with $\nu_0=(\beta \nu +\beta -1)/\beta$ and  $Y \to Y/2$.

The spectral statistics of WBE is discussed in detail \cite{sp}; the above mapping can then be used to determine the  singular value statistics  of $\rho_c(C)$ (described by eqs.(\ref{amn2}, \ref{amn3})) and more details (beyond section VI. B, C) about the spectral statistics of $\rho(H)$ (described  by eqs.(\ref{amn2a}, \ref{amn3a})). Similarly  the  statistics of the eigenfunction $X_n$ of $L_1$, with $\rho_{L_1}$ described by a WBE,  is discussed  in \cite{pslg}  and the steps can be generalized to the eigenfunction $Z_n$ of $L_2$;  this information can further be used to  derive the statistics of the eigenfunctions  $U_n$ of $\rho(H)$ corresponding to non-zero modes.

\subsection{Multi-parametric Gaussian Ensembles}

The results and insights mentioned in previous section have another important application i.e in the domain of multi-parametric ensembles of chiral Hermitian, Wishart and non-chiral Hermitian matrices. This can be elucidated as follows.

Consider an ensemble $\rho_m(H)=\rho_m(C) F_c F_h$ of chiral Hermitian matrices with $H$ still given by eq.(\ref{ch1}) but ensemble density $\rho_m(C)$  of $C$-matrices now described by 
\begin{eqnarray}
\rho_m(C)=\mathcal{N} {\rm e}^{-\sum_{k,l; s} {1\over 2 h_{kl;s}} (H_{k,N+l;s} -b_{kl;s})^2} \; F_c \; F_h
\label{rhomc}
\end{eqnarray}
with $\sum_{k,l,s} \equiv \sum_{k,l=1}^N \sum_{s=1}^{\beta}$ and $F_c, F_h$ same as given below eq.(\ref{rhoh}).  As discussed in \cite{psmulti}, the diffusion equation for $\rho_m(H)$   is analogous to eq.(\ref{chi8}) but $Y$ now corresponds to a function of all ensemble parameters
$Y=-\frac{1}{2 M\gamma} \; {\rm ln} \left[\prod_{k,l;s}|1- 2 \; \gamma  \; h_{kl;s} | \;|b_{kl;s}|^2\right]+{\rm const}$ (see section II.B of \cite{psmulti}).  

Although governed by different parameters, the analogy of diffusion equations for  $\rho(H)$ and $\rho_m(H)$ implies similar analogies for the diffusion of their eigenvalues and eigenfunctions. Consequently eqs.(\ref{f8}-\ref{chx20}) are expected to be applicable for the ensemble (\ref{rhomc}) too; this is indeed confirmed  by the analogy of spectral diffusion equation of $\rho(H)$ (eq.(\ref{ch19}))  with  that for  $\rho_m(H)$ (see eq.(9) derived in \cite{psmulti} by an alternative route).  The spectral and strength statistics of multi-parametric chiral ensembles can then be mapped to single parametric chiral Brownian ensembles;   the numerical study discussed in \cite{psmulti} verifies the above claim for a prototypical spectral fluctuation measure, namely, nearest neighbor spacing ratio distribution.

Following the analogy and discussion in section VII.C,  the local spectral statistics (away from the zero energy) of a multi-parametric chiral  ensemble can also be mapped to that of a non-chiral Hermitian Brownian ensemble and thereby to latter's multi-parametric counterparts \cite{ps-all}. Furthermore, the discussion in section VII A and B  indicate that  the local statistics of a multi-parametric chiral  ensemble is also connected to generalized CS Hamiltonian and Wishart Brownian ensembles.

\subsection{Bipartite Disordered Lattices}

Fundamental as well as technological interest in chiral disordered systems has motivated many studies of their statistical behavior  with intense focus on the  density of states  and eigenfunctions near zero energy \cite{gade, gw, mdh, ek}. Many of these studies are based on the bipartite disordered lattices (BDL) which can be  modeled by eq.(\ref{rhomc}) (see \cite{psmulti} by examples) and therefore their statistical properties can be mapped to Ch-BE through complexity parametric formulation (CPF). The results obtained by previous studies \cite{gade, gw, mdh, ek} are therefore expected to be consistent with corresponding results for Ch-BE. To indicate that this indeed seems to be the case, here we give some examples:

(i) For a BDL, the chiral state i.e the eigenstate corresponding to $e=0$ is non-zero in one sub-lattice only \cite{ek}. From eq.(\ref{uz}), this is the case for Ch-BE too.  

(ii) The chiral state for a $d$-dimensional BDL turns out to be neither localized nor extended but critical for all dimensions and sensitive to boundary conditions \cite{ek,ek1,ita}.  As discussed below eq.(\ref{uzp4}), a similar behavior is expected for the chiral state of Ch-BE too. 

(iii) In absence of time-reversal symmetry, the density of states near $e \sim 0$ for a BDL ($d >1$) given in \cite{gade} is analogous to eq.(\ref{rr4}) with $b=1/2, \eta=1$. A similar study \cite{mdh}  for another BDL (with $d >1$) in presence of time-reversal symmetry  gives $b=2/3, \eta=1$. As mentioned near eq.(\ref{rr4}), both these results agree only approximately with our theory.  The result of \cite{lfsg} indicating a power law divergence of density near $e=0$ however is consistent with our formulation.

(iv) As reported in \cite{ek}, the level statistics for the chiral state  $(e=0)$ of an infinite size BDL is  intermediate between Poisson and Wigner-Dyson statistics. Following eq.(\ref{rn}), an analogous behavior is predicted for Ch-BE as well as for eq.(\ref{rhomc}) if $\Lambda_e( e\sim 0)$ is size-independent (see \cite{psmulti} for details);  the analogy is also verified numerically in \cite{psmulti}.

(v) The study \cite{ek} indicates that the local spectral fluctuations in regions $e\not=0$ for a BDL are expected to be analogous to those of non-chiral case. As 
indicated near eq.(\ref{rn1}), this is also the case for Ch-BE and thereby for eq.(\ref{rhomc}).

(v) As discussed in \cite{psmulti} (also see \cite{psand} in context of non-chiral Anderson Hamiltonian), the dimensionality dependence in the complexity parameter $\Lambda_e$ (sole governor of the the spectral fluctuation measures besides global constraints) enters through $Y$ as well as the local mean level spacing $\Delta_e$. The critical statistics of eq.(\ref{rhomc}) therefore occurs for system conditions which results in $\Lambda_e$ as size-independent.  The observed dimensionality dependence of the critical statistics of non-chiral state in BDL is therefore again in agreement with our prediction.

\section{Conclusion}

In the end, we summarize with a brief discussion of our main results and open questions.

Based on an exact response of the eigenvalues and eigenfunctions of a chiral matrix to a chirality preserving perturbation, we derive diffusion equations for their joint probability density functions. The information is then used to analyze the statistical behavior near zero energy. Following complexity parameter formulation discussed in \cite{psmulti}, our results are also applicable to a wide range of complex systems, modeled by multi-parametric Gaussian ensembles with chiral symmetry e.g. disordered bipartite lattices.  As expected, previously known results for the density of states and localization behavior, obtained by system specific approaches,  are indeed consistent with ours and thereby reconfirms the complexity parametric formulation of the statistical properties.

Another important insight, our work provides, is about the deep web of connections underlying the world of chiral complex systems.
Previous studies of many non-stationary ensemble e.g.  basis-dependent Hermitian non-chiral ensembles, circular ensembles, Wishart ensemble etc. indicate the standard Calogero-Sutherland Hamiltonian (CSH) (and its variants) as the generator of the dynamics of their eigenvalues; this led to  CSH referred as the ''universal Hamiltonian'' \cite{pcalo, sp}.
With help of complexity parameter formulation, here we establish a similar connection 
with a generalized version of CSH which further lends credence to the idea of CSH 
as the underlying universal hidden structure governing the dynamics of the eigenvalues 
of complex systems. The appearance of CS Hamiltonian is not restricted only to the spectral properties; it has been known to manifest itself in other properties of complex systems too \cite{pcalo}. A detailed investigation of the CS hamiltonian in arbitrary dimension can therefore give a lot of
useful information about variety of complex systems and is very much
desirable.

Our study still leaves many questions unanswered. The first and foremost among them is the solutions of various diffusion equations for the spectral and strength jpdf derived here and an explicit formulation of the fluctuations measures.  Another important question is about the  transition from chiral ensembles to non-chiral ensembles as chiral symmetry is partially broken.  Although this has been discussed in context of chiral ensemble appearing in QCD (e.g. see \cite{kk}),  the information in case of other complex systems e.g. bipartite lattices is still missing.

.....



\appendix

\section{Derivation of eqs.(\ref{ch9}, \ref{ch10})}

The eigenvalue equation $H U= U E $ or $E= U^{\dagger} H U$, with $U$ and $E$ as the eigenvector and eigenvalue matrix of $H$, leads to following: 
     
\begin{eqnarray}
 \sum_{j=1}^{N_{\nu}} H_{ij} U_{jn} = e_n U_{in} 
 \label{g1}
 \end{eqnarray}
where $H_{ij}=H_{ij;1} + i H_{ij;2}$. Note here $H_{ij}=0$ for  $i,j=1 \le N$ or $i,j >N$. 
With $H$ as a $N_{\nu} \times N_{\nu}$ Hermitian matrix, $U$ is a unitary matrix and $E$ is real 
diagonal. Although the additional constraint of chirality leads to equal and opposite pairs or zero eigenvalues and various relations among their eigenfunctions (section II), this does not affect the derivation of  results for ${\partial e_n \over\partial H_{kl;s}}$ and ${\partial U_{rn} \over\partial H_{kl;s}}$ with $e_n, U_n$ as non-zero energy state can still be expressed in the same form as 
for a Hermitian matrix without chirality; the derivation for the latter is discussed in \cite{pswf, ps-all}. The results for the zero energy i.e chiral state however can not be derived by the same route. 

 Differentiating both sides of the above equation with respect to $H_{k,N+l;s}$ (with $k \le N, N < l \le N+\nu$ and $s=1\; {\rm or}\; 2$ ), we get
\begin{eqnarray}
\sum_j {\partial U_{jn}\over\partial H_{k,N+l;s}} H_{i,N+j} + 
\sum_j U_{jn} {\partial H_{ij}\over\partial H_{k,N+l;s}}  &=& 
e_n {\partial U_{in}\over\partial H_{k,N+l;s}}  +  
{\partial e_n \over\partial H_{k,N+l;s}} U_{in}  
\label{g2}
\end{eqnarray}

Now as $\sum_i U_{in}^* U_{im} = \delta_{nm}$, 
multiplying both the sides by $U_{in}^{*}$  followed by a 
summation over all $i$'s leads to

\begin{eqnarray}
 {\partial e_n \over\partial H_{k,N+l;s}} &=& 
\sum_{i,j} U_{in}^{*} {\partial H_{ij}\over\partial H_{k,N+l;s}} U_{jn}
\label{g3}
\end{eqnarray}

which further gives, using $H_{ij}=H_{ji}^*$,

\begin{eqnarray}
 {\partial e_n \over\partial H_{k,N+l;s}} &=& 
 i^{s-1}
 \left[ U_{N+l,n} U_{kn}^{*} + (-1)^{s+1} \; U_{N+l,n}^{*} U_{kn}\right]
 \label{g4}
\end{eqnarray}

The first order variation of the eigenfunction component can also be 
derived from eq.(\ref{g2}) but it depends on whether it corresponds to 
a chiral or non-chial state. Multiplying both the sides of eq.(A2) by $U_{im}^{*}$ ($m \not=n$) followed by a summation over all $i$'s, we get 
\begin{eqnarray}
\sum_j U_{jm}^{*} {\partial U_{jn}\over\partial H_{k,N+l;s}} &=& 
{ i^{s-1} \over {e_n -e_m}}\sum_{i,j} U_{im}^{*} {\partial H_{ij}\over\partial H_{k,N+l;s}} U_{jn} \nonumber \\
 &=& {i^{s-1}\over {e_n-e_m}} \left( U_{km}^{*} U_{N+l,n} +(-1)^{s+1} U_{N+l,m}^{*} U_{kn} \right)
 \label{g5}
\end{eqnarray}

{\bf Case for non-zero eigenvalue:}
A multiplication of both the sides by $U_{rm}$ followed by a summation 
over all $m=1 \to 2N+\nu$ then gives
\begin{eqnarray}
 {\partial U_{rn}\over\partial H_{k,N+l;s}} &=& i^{s-1} 
\sum_{m\not=n} {U_{rm}\over {e_n -e_m}}
 \left( U_{km}^{*} U_{N+l,n} +(-1)^{s+1} U_{N+l,m}^{*} U_{kn} \right)
 \label{g6}
\end{eqnarray}

\vspace{0.1in}

{\bf Case for zero eigenvalue:} 
As right side of eq.(\ref{g5}) now diverges at $e_m=0$, it can no longer be summed over $\sum_{m=1}^{2N+\nu} U_{rm}$. Thus multiplying both sides by $U_{rm}$ and summing over ${m=1 \to 2N}$, we have 
\begin{eqnarray}
{\partial U_{rn}\over\partial H_{k,N+l;s}} - \sum_{m=2N+1}^{2N+\nu} \sum_j U_{rm} U_{jm}^{*} {\partial U_{jn}\over\partial H_{k,N+l;s}} 
&=&  X_r^{(n)} 
\label{g7}
\end{eqnarray}
with $X_r^{(n)} = -i^{s-1} \sum_{m=1}^{2N} {U_{rm} \over {e_m}}
 \left( U_{km}^{*} U_{N+l,n} +(-1)^{s+1} U_{N+l,m}^{*} U_{kn} \right)$.
 The above can be rewritten as 
\begin{eqnarray}
\sum_{j=1}^{2N+\nu} T_{rj} \; {\partial U_{jn}\over\partial H_{kl;s}} &=&  X_r^{(n)} 
 \label{g8}
\end{eqnarray}
where $T_{rj}$ and $X_r^{(n)}$ are given by eqs.(\ref{ch11}, \ref{ch12}). The above set of equations can also be written as a  matrix equation $T. \left[\partial U_{rn}\right]= \left[ X_r^{(n)}\right] $ (with notation $\partial U_{rn} \equiv {\partial U_{rn}\over\partial H_{kl;s}}$). An inversion of the above equation then leads to eq.(\ref{ch10}).

\section{{Derivation of eqs.(\ref{ch13}, \ref{ch15}, \ref{ch12a})} }

Multiplying eq.(\ref{g4}) by $H_{kl;s}$, summing over $k=1 \to N, l=N+1 \to N+\nu$ and $s=1 \to \beta$
and subsequently using relations, $\sum_s i^{s-1} H_{kl;s}=H_{kl}$,  $\sum_s i^{s-1} (-1)^{s+1} H_{kl;s} =H_{kl}^* = H_{lk}$ and $E= U^{\dagger} H U$ gives 
\begin{eqnarray}
\sum_{k,l;s=1,N+1;1}^{N,N+\nu, \beta}  {\partial e_n \over\partial H_{kl;s}} H_{kl;s}  
&=& \sum_{k,l=1,N+1}^{N,N+\nu}  \left( H_{kl} U_{ln} U_{kn}^{*} 
 + H_{kl}^* U_{ln}^{*} U_{kn} \right) \label{gg9} \\
&=& \sum_{k, l=1}^{2N+\nu}  H_{kl} U_{ln} U_{kn}^{*} = e_n 
 \label{g9}
\end{eqnarray}
with eq.(\ref{g9}) results  by interchanging $k,l$ in the $2nd$ term of of eq.(\ref{gg9}).

A summation over a product of eq.(\ref{g4}) for two different energies, say $e_n, e_m$ and subsequently using unitary nature of $U$ gives 
\begin{eqnarray}
\sum_{k,l;s=1,N+1;1}^{N,N+\nu, \beta}
 {\partial e_n \over\partial H_{kl;s}}   
  {\partial e_m \over\partial H_{kl;s}}   
&=& \sum_{k, l=1}^{2N+\nu}  
\left[ U_{ln} U_{kn}^{*} U_{km} U_{lm}^* + 
 U_{ln}^{*} U_{kn} U_{lm} U_{km}^* \right] \label{gg10}\\
&=& 2\; \delta_{mn}
\label{g10}
\end{eqnarray}
with last equality follows from  unitary nature of $U$.

Similarly a  summation over a product of eq.(\ref{g4}) with eq.(\ref{g6}) and subsequently using the relation $U^{\dagger}U= 1$ gives 
\begin{eqnarray}
\sum_{k,l;s=1,N+1;1}^{N,N+\nu, \beta} {\partial e_i \over\partial H_{kl;s}}{\partial U_{nj} \over\partial H_{kl;s}} &=&  0 
 \end{eqnarray}

Differentiating eq.(\ref{g4}) with respect to $H_{kl;s}$, subsequently using eq.(\ref{g6}) followed by a summation  now leads to

\begin{eqnarray}
\sum_{k,l;s=1,N+1;1}^{N,N+\nu, \beta}  {\partial^2 e_n \over\partial H_{kl;s}^2}   
&=& 2 \; \beta \;  \sum_{k,l=1,N+1}^{N,N+\nu} \sum_m {1\over e_n -e_m} 
\left[ U_{km} U_{km}^* U_{ln} U_{ln}^* 
+ U_{kn} U_{kn}^* U_{lm} U_{lm}^* \right]\\  
&=& 2 \; \beta \; \sum_{k, l=1}^{2N+\nu}  \sum_m {1\over e_n - e_m} 
\left[ U_{km} U_{km}^* U_{ln} U_{ln}^* \right]
\label{g12}
\end{eqnarray}
Now by using the unitary relation $\sum_j U_{jm}^* U_{jm} = 1$, 
one  obtains the desired relation (7).  

\section{Derivation of eq.(\ref{chh16})}

Differentiating eq.(\ref{g5}) with respect to $H_{kl;s}$ for $m \le 2N, n > 2N$, summing over $k,l,s$  and then using eq.(\ref{chh18}), we have 
\begin{eqnarray}
\sum_{j=1}^{2N+\nu}  \sum_{k,l,s} U_{jm}^* {\partial^2 U_{jn} \over\partial H_{kl;s}^2} = 
\sum_{k,l,s}  {\partial f_{klmn;s}\over\partial H_{kl;s}}  
 \end{eqnarray}
where 
\begin{eqnarray}
f_{klmn;s} \equiv {-i^{s-1} \over {e_m}}\left( U_{km}^{*} U_{ln} +(-1)^{s+1} U_{lm}^{*} U_{kn} \right). 
\label{fmn}
\end{eqnarray}

Now Multiplying the above equation  by $U_{rm}$ ($m \not=n$) followed by a summation over all $m=1 \to 2N$ and writing $V_{rj}=\sum_{m=1}^{2N} U_{rm} U_{jm}^*$, we get 
\begin{eqnarray}
V. \left[\partial^2 U_{rn}\right]= \left[G_r^{(n)}\right] 
\end{eqnarray}
where $G_r^{(n)}=\sum_{m=1}^{2N}  \sum_{k,l,s} U_{rm}  {\partial f_{klmn;s}\over\partial H_{kl;s}} $. An inversion of the above equation then leads to $\left[\partial^2 U_{rn}\right]= V^{-1} \left[G_r^{(n)}\right] $ and thereby
\begin{eqnarray}
\sum_{kls} \partial^2 U_{rn} = \sum_a (V^{-1})_{ra} G_a^{(n)} = \sum_a  \sum_{m=1}^{2N}  \sum_{k,l,s} (V^{-1})_{ra} U_{rm}  {\partial f_{klmn;s}\over\partial H_{kl;s}}
\label{fmn2}
\end{eqnarray}
Now differentiating eq.(\ref{fmn}) with respect to $H_{kl;s}$ gives a sum over products of eigenfunction components with derivatives of $e_n$ and $U_{jn}$. Using eq.(\ref{g4}, \ref{g6})   
 and a summation over indices $k,l,s$, this leads to $\sum_{k,l,s} {\partial f_{klmn;s}\over\partial H_{kl;s}}=0$ and thereby $\sum_{kls} {\partial^2 U_{rn} \over\partial H_{kl;s}^2} =0$ for $n > 2N$.

\section{Derivation of eq.(\ref{chh19})}

A differntiation of eq.(\ref{peef1}) with respect to Y gives 
\begin{eqnarray}
{\partial P_{ef} \over \partial Y} = \int  \left[{\partial P^{(c)}_{ef,ev}  \over \partial Y} \right]\; {\rm D}e
\label{peef2}
\end{eqnarray} 
with ${\rm D}e \equiv \prod_{n=1}^{2N} {\rm d}e_n$. 
Substitution of eq.(\ref{f8}) again on the right side and subsequently using eq.(\ref{peev1})  now gives 
\begin{eqnarray}
{\partial P_{ef} \over \partial Y} = ({\mathcal L}_I+  {\mathcal L}_I^*) P^{(c)}_{ef,ev}  + \int  {\mathcal L}_E P_{ef,ev}  \; {\rm D}e.
\label{peef3}
\end{eqnarray} 
with $ {\mathcal L}_I$ and $ {\mathcal L}_E$ given by eq.(\ref{chx20}) and eq.(\ref{dev}). 
Further eq.(\ref{f9}) leads to
\begin{eqnarray}
 \int  \left[ {\mathcal L}_E  P^{(c)}_{ef,ev}  \right]\; {\rm D}e  &=& \sum_{n=1}^{2N+\nu} K_{n}
\label{peef4}
 \end{eqnarray}
 where
\begin{eqnarray}
K_n &=& \int  \frac{\partial}{\partial e_n}\left[\frac{\partial}{\partial e_n}- \sum_{m=1}^N \frac{\beta }{e_n- e_m} + {\gamma\over v^2} \; e_n\right]
\; P^{(c)}_{ef,ev} \; {\rm D}e  \nonumber \\
&=&  \left[ \left(1- \sum_{m=1}^N \frac{\beta }{e_n- e_m} + {\gamma\over v^2} \; e_n\right) \; P^{(c)}_{ef,ev}\right] _{e_1,\ldots, e_N \to \infty}
 \label{peef5}
 \end{eqnarray}
With $P^{(c)}_{ef,ev}(\{e_n \}, \{U_{rn}\}) \to 0$ as $\{e_n\} \equiv (e_1,\ldots, e_N) \to \infty$ , we have $K_n=0$ and thereby $ \int  \left[ {\mathcal L}_E  P^{(c)}_{ef,ev}  \right]\; {\rm D}e \to 0$. Eq.(\ref{peef3}) then reduces to eq.(\ref{chh19}).

\section{Derivation of eq.(\ref{ch19})}

A differentiation of eq.(\ref{peev1}) with respect to Y gives 
\begin{eqnarray}
{\partial P_{ev} \over \partial Y} = \int  \left[{\partial P^{(c)}_{ef,ev}\over \partial Y}  \right]\; {\rm D}U
\label{peev2}
\end{eqnarray} 
with ${\rm D}U \equiv \prod_{k,n=1}^{2N+\nu} {\rm d}U_{kn}$. 
Substitution of eq.(\ref{f8}) on the right side and again using eq.(\ref{peev1})  now gives 
\begin{eqnarray}
{\partial P_{ev} \over \partial Y} = L_E P_{ev} +{\beta^2 \over 8} \int (  {\mathcal L}_U +  {\mathcal L}_U^*) P^{(c)}_{ef,ev}  \;  {\rm D}U.
\label{peev3}
\end{eqnarray} 
with $ {\mathcal L}_U$ and $ {\mathcal L}_E$  given by eq.(\ref{f9}) and eq.(\ref{f9p}). 
Further eq.(\ref{f9}) leads to
\begin{eqnarray}
 \int  \left[ {\mathcal L}_U  P^{(c)}_{ef,ev}  \right]\; {\rm D}U  &=& \sum_{k,l,j,n} \; L_{jnkl} + \sum_{k,j,n,m \atop m\not=n}   
 \; M_{kjnm} + \sum_{j,n,m=1 \atop m\not=n}  N_{jnm} 
 \label{peev4}
 \end{eqnarray}
where 
\begin{eqnarray} 
L_{jnkl} &=& \int  {\partial^2 \over \partial U_{jn} \partial U_{kl}}
\left( { S_{jlkn} \; P^{(c)}_{ef,ev}  \over (e_l-e_n)^2} \right) \; {\rm D}U  \\
M_{kjnm} &=& \int {\partial^2 \over \partial U_{jn} \partial U_{kn}^*} 
\left({ D_{jmkm} \; P^{(c)}_{ef,ev}  \over (e_n-e_m)^2} \right) \; {\rm D}U \\
N_{jnm} &=& \int {\partial \over \partial U_{jn} }  
\left({ U_{jn}  \; P^{(c)}_{ef,ev}  \over (e_n-e_m)^2} \right) \; {\rm D}U
\label{peev5}
\end{eqnarray} 
An integration over $U_{jn}$ and $U_{kl}$ of the terms inside square bracket above further gives
\begin{eqnarray}
 L_{jnkl} &=&
    \left(  { S_{jlkn} \; P^{(c)}_{ef,ev}  \over (e_l-e_n)^2} \right)_{\{U_{rn}\}=1}  - \left( { S_{jlkn} \; P^{(c)}_{ef,ev}  \over (e_l-e_n)^2} \right)_{\{U_{rn}\}=-1}. 
 \label{peev6}
\end{eqnarray} 
With $P^{(c)}_{ef,ev}(\{e_n \}, \{U_{rn}\}) \to 0$ as $\{U_{rn}\} \equiv (U_{11}, U_{12},\ldots, U_{2N+\nu, 2N+\nu}) \to \pm 1$, we have $I_{jnkl} =0$. Following similar reasoning also gives $M_{kjnm} =0$ and $N_{jnm}=0$.
These on substitution in eq.(\ref{peev4}) leads to $\int  \left[ {\mathcal L}_U  P_{ef,ev}  \right]\; {\rm D}U =0$. Proceeding similarly, it can be shown that $\int  \left[ {\mathcal L}_U^*  P_{ef,ev}  \right]\; {\rm D}U =0$. Eq.(\ref{peev3}) now gives ${\partial P_{ev} \over \partial Y} = L_E P_{ev}$ . The latter along with eq.(\ref{dev}) and eq.(\ref{pe1}) now leads to eq.(\ref{ch19}).

\section{Derivation of eq.(\ref{ptc1}) and eq.(\ref{ptc2})}
As eq.(\ref{uzp4}) is first order in $\Lambda_t$, It is appropriate to consider following  solution:
$P_t(t;\Lambda_t) = {\rm exp}(- E \Lambda_t) T(t)$. 
 where $E$  is an arbitrary constant. Substitution of the above expression for $P_t$ in eq.(\ref{uzp4}) 
leads to 
\begin{eqnarray}
   t \; {\partial^2 T \over \partial t^2} +   {\partial T \over \partial t} + E T= 0.
\label{v2}
\end{eqnarray}
Clearly, for $P_t$ to remain finite and non-zero as $\Lambda_t \to \infty$, we must have $E=0$. The solution of eq.(\ref{v2}) for the latter case can be given as $T(t)=C_1+C_2 \log t$ with $C_1, C_2$ as constants determined by bondary conditions on $P_t$; as expected, the solution is same as that of eq.(\ref{uzp4}) with  ${\partial P_t \over \partial \Lambda_t}=0$.

The general solution of eq.(\ref{v2}) for $E > 0$ can be given as 
\begin{eqnarray}
T(t; \Lambda_t, E) = C_1 \; J_0(2\sqrt{E t})  + 2 \;  C_2 \; N_0(2\sqrt{E t}) 
\label{v3}
\end{eqnarray}
with  $J_0$ and $N_0$ as the Bessel functions of order zero of the first and second kind, and, $C_1, C_2$ as the constants of integration. As $N_0(x) \to -\infty$ as $x \to 0$, clearly $C_2=0$ for cases with $P(t,Y)$ finite for $0 \le t \le 1$.  The  solution  can then be given as 
\begin{eqnarray}
P_t(t; \Lambda_t) = C_1 \;  {\rm exp}(- E \Lambda_t) \; J_0(2\sqrt{E t}),  \label{v4}
\end{eqnarray}
To determine $C_1$, we now use the boundary condition at $t=1$: $P(1, \Lambda_t) \to 0$, the latter implies $J_0(2\sqrt{E t})=0$.  Referring $z_n$ as the $nth$ zero of $J_0(z)$ with $z_1 < z_2 \ldots z_n$, this condition  can then be satisfied for  $E=z_n^2/4$ for $n=1 \to \infty$. The general solution can then be written as 
\begin{eqnarray}
P_t(t; \Lambda_t) = \sum_{n=1}^{\infty} B_n \;  {\rm exp}(- z_n^2 \Lambda_t/4) \; J_0(z_n\sqrt{ t}),  
\label{v6}
\end{eqnarray}
The constants $B_n$ can be determined from the initial condition 
$P_t(t;0)=P_0(t) =\sum_n B_n \;  J_0(z_n\sqrt{ t})$ along with orthogonality relation of $J_0$:
$\int_0^1 t J_0(z_n t) J_0(z_m t) {\rm d}t = J_1^2(z_n) \delta_{nm}$. 
This gives $B_n$ in terms of initial distribution $P_t(t;0)$
\begin{eqnarray}
B_n =  {2 \over  J_1^2(z_n)} \; \int_0^1 t \; J_0(z_m t) \; P_t(t; 0) \;  {\rm d}t
\label{bnt}
\end{eqnarray}

Substitution of the above in eq.(\ref{v6}) gives $P_t(t; \Lambda_t)$ in terms of $P_t(t; 0)$. Although eq.(\ref{v6})  contains a sum over infinite number of terms but $J_0(z_n t)$ decays rapidly with increasing $z_n$. This along with term  ${\rm exp}(- z_n^2 \Lambda_t/4)$ renders the contribution from higher order terms insignificant and only first few of them are sufficient to evaluate $P_t(t; \Lambda_t)$ for any initial condition. The first five zeros of $J_0$ can be given as 
$z_1=2.40483, z_2=5.52008, z_3=8.65373, z_4=11.79153, z_5=18.07106$.

Let us now consider the cases with diverging $P_t(t, Y)$  at $t \to 0$. Clearly $C_1=0$ for these cases. The solution at $t\to 0$ now becomes
\begin{eqnarray}
P_t(t; \Lambda_t) = C_2 \; {\rm exp}(- E \Lambda_t) \; N_0(2\sqrt{E t}) 
\label{v5}
\end{eqnarray}
As for $C_1$, $C_2$ can also be determined by the boundary condition at $t=1$. With 
$P(1, \Lambda_t) \to 0$, the latter implies  $N_0(2\sqrt{E t})=0$ for the 2nd case. 
Referring $z_n$ now as the $nth$ zero of $N_0(z)$ with $z_1 < z_2 \ldots z_n$, the condition $P(1, \Lambda_t) \to 0$ can be satisfied for  $E=z_n^2/4$ for $n=1 \to \infty$. The general solution can then be written as 
\begin{eqnarray}
P_t(t; \Lambda_t) = \sum_{n=1}^{\infty} B_n \;  {\rm exp}(- z_n^2 \Lambda_t/4) \; N_0(z_n\sqrt{ t}),  
\label{v6}
\end{eqnarray}
Due to lack of orthogonality relations for $N_0$,  determination of the constants $B_n$  however is no longer straightforward. With a known initial distribution $P_t(t,0)$ and its moments, 
one possible route is to express $B_n$ in terms of the moments at $\Lambda_t=0$. As 
$P_t(t;0) =\sum_n B_n \;  N_0(z_n\sqrt{ t})$, this in turn leads to  an infinite set of equations 
$\langle t^m \rangle_{\Lambda_t=0} =\sum_n B_n \; g_{mn}$ where 
$g_{mn} = \int_0^1 t_n \;  N_0(z_n\sqrt{ t}) {\rm d}t$.  This leads to relation
$B= G^{-1} \Gamma$ where $B \equiv \left[B_n \right]$ and $\Gamma \equiv \left[ \langle t^m \rangle_{\Lambda_t=0} \right]$ are column vectors and $G \equiv \left[ g_{mn} \right]$ is a matrix of infinite size. Here again, as in the previous case, these matrices can be truncated to finite sizes without significantly affecting the result.

\end{document}